\documentclass[10pt,journal,compsoc]{IEEEtran}

\ifCLASSOPTIONcompsoc
  \usepackage[nocompress]{cite}
\else
  \usepackage{cite}
\fi

\AtBeginDocument{%
  \providecommand\BibTeX{{%
    \normalfont B\kern-0.5em{\scshape i\kern-0.25em b}\kern-0.8em\TeX}}}

\usepackage{cite}
\usepackage{amsmath,amssymb,amsfonts}
\usepackage{algorithmic}
\usepackage{graphicx} 
\usepackage{booktabs}
\usepackage{textcomp}
\usepackage{subcaption}
\usepackage{tcolorbox}
\usepackage{amsmath}
\usepackage{amssymb}
\usepackage{enumerate}
\usepackage{enumitem}
\usepackage{alltt}
\usepackage{url}
\usepackage{caption}
\usepackage{subcaption}
\usepackage{xspace}
\usepackage{hyperref}
\usepackage[export]{adjustbox}
\usepackage{algorithm}
\usepackage{balance}

\usepackage{listings}
\lstdefinelanguage{diff}{
  language=Java,
  morecomment=[f][\color{blue}]{@@},     
  morecomment=[f][\color{red!60!black}]-,         
  morecomment=[f][\color{green!60!black}]+,       
  morecomment=[f][\color{magenta}]{---}, 
  morecomment=[f][\color{magenta}]{+++},
}
\usepackage{stackengine}
\usepackage{boxhandler}
\usepackage{soulutf8} 

\newcommand{\changing}[1]{{\color{black} #1}}

\newboolean{showcomments}
\setboolean{showcomments}{false} 
\ifthenelse{\boolean{showcomments}}
 { \newcommand{\mynote}[2]{
      \fbox{\bfseries\sffamily\scriptsize#1}
        {\small$\blacktriangleright${{{#2}\bf }}$\blacktriangleleft$}}}
        { \newcommand{\mynote}[2]{}}

\newcommand{\OurTool}{\text{$\mu$\textsc{Bert}}\xspace} 
\newcommand{\Toolconv}{\text{$\mu$\textsc{Bert}$_{conv}$}\xspace} 
\begin{document}

 \title{Efficient Mutation Testing via Pre-Trained Language Models}
\author{
\IEEEauthorblockN{Ahmed Khanfir}
,
\and
\IEEEauthorblockN{Renzo Degiovanni}
,
\and
\IEEEauthorblockN{Mike Papadakis}
and
\and
\IEEEauthorblockN{Yves Le Traon}\\
\IEEEauthorblockA{SnT, University of Luxembourg, Luxembourg}
\IEEEcompsocitemizethanks{\IEEEcompsocthanksitem A. Khanfir, R. Degiovanni, M. Papadakis, Y. Le Traon are with the University of Luxembourg, Luxembourg.}}


\IEEEtitleabstractindextext{
\begin{abstract}
Mutation testing is an established fault-based testing technique. It operates by seeding faults into the programs under test and asking developers to write tests that reveal these faults. These tests have the potential to reveal a large number of faults -- those that couple with the seeded ones -- and thus are deemed important. To this end, mutation testing should seed faults that are both ``natural'' in a sense easily understood by developers and strong (have high chances to reveal faults). To achieve this we propose using pre-trained generative language models (i.e. CodeBERT) that have the ability to produce developer-like code that operates similarly, but not exactly, as the target code. This means that the models have the ability to seed natural faults, thereby offering opportunities to perform mutation testing. We realise this idea by implementing {\OurTool}, a mutation testing technique that performs mutation testing using CodeBert and empirically evaluated it using \changing{689} faulty program versions. Our results show that the fault revelation ability of \OurTool is higher than that of a state-of-the-art mutation testing (PiTest), yielding tests that have up to \changing{17\%} higher fault detection potential than that of PiTest. Moreover, we observe that {\OurTool} can complement PiTest, being able to detect \changing{47} bugs missed by PiTest, while at the same time, PiTest can find \changing{13} bugs missed by {\OurTool}. 

\end{abstract}

\begin{IEEEkeywords}
Fault Injection, Mutation Testing, Pre-Trained Language Models
\end{IEEEkeywords}
}

\maketitle    

\section{Introduction}

Mutation testing aims at seeding faults using simple syntactic transformations~\cite{DeMilloLS78}. 
These transformations, also known as mutation operators are typically constructed based on syntactic rules crafted based on the grammar of the target programming language~\cite{0020331}, i.e. replacing an arithmetic operator with another such as a \texttt{+} by a \texttt{-}. 
Unfortunately, such techniques generate mutants (seeded faults), many of which are ``unatural'', i.e., non-conforming to the way developers code, 
thereby perceived as unrealistic by developers~\cite{BellerWBSM0021}. At the same time, the syntactic-based fault seeding fails to capture the semantics of the code snippets that they apply, leading to numerous trivial or low utility faults \cite{PapadakisCT18}. 


To deal with the above issue we propose forming natural mutations by using big code. Thus, we aim at introducing modifications that follow the implicit rules, norms and coding conventions followed by programmers, by leveraging the capabilities of pre-trained language models to capture the underlying distribution of code and its writing, as learned by the pre-training process on big code.  

To this end, we rely on CodeBERT~\cite{DBLP:conf/emnlp/FengGTDFGS0LJZ20}, an NL-PL bimodal language model that has been trained on over 6.4 million programs. More precisely, we use its Masking Modelling Language (MLM) functionality, which given a code sequence with a masked token, predicts alternative replacements to that token, that is best matching the sequence context.
This is important, since the predictions do not follow fixed predefined patterns as is the case of conventional mutation testing, but are instead adapted to fit best the target code.  
For instance, given a sequence \texttt{int a = 1;}, we pass a masked version of it as \texttt{int a = <mask>;}, then CodeBERT by default proposes 5 predictions sorted by likelihood score: \texttt{0}, \texttt{1}, \texttt{b}, \texttt{2},  and \texttt{10}.
Being the most likely fitting tokens to the code context, our intuition is that replacing the masked token with these predictions would induce ``natural'' mutants. 

Precisely, we introduce {\OurTool}, a mutation testing approach that uses a pre-trained language model (CodeBERT) to generate mutants by masking and replacing tokens with the aim of forming natural mutants. 
{\OurTool} iterates through the program statements and modifies their token. In particular, {\OurTool} proceeds as follows: (1) it selects and masks one token at a time
; (2) feeds CodeBERT with the masked sequence and obtains the predictions; (3) creates mutants by replacing the masked token with the predicted ones; and (4) discards non-compilable, duplicate and equivalent mutants (mutants syntactically equal to original code). 

Recent research~\cite{khanfir2020ibir} has shown that some real faults are only captured by using complex patterns, i.e. patterns that require more than one token mutation. To account for such cases, {\OurTool} is equipped with additive mutations,  i.e., mutations that add code (instead of deleting or altering). For example, consider a boolean expression $\texttt{e}_1$ (typically present in \texttt{if}, \texttt{do}, \texttt{while} and \texttt{return} statements), which is mutated by {\OurTool} by adding a new condition $\texttt{e}_2$
, thereby generating a new condition $\texttt{e}_1 \texttt{||} \texttt{e}_2$ (or $\texttt{e}_1 \texttt{\&\&} \texttt{e}_2$), which is then masked and completed by CodeBERT. For instance, given a condition \texttt{if(a == b)}, {\OurTool} produces a new condition \texttt{if(a == b || a > 0)} that is masked and produces \texttt{if(a == b || b > 0)}.

We implement \OurTool, and evaluate its ability to serve the main purposes of mutation testing, i.e. guiding the testing towards finding faults. We thus, evaluate it using \changing{689} faults from Defects4J and asses \OurTool effectiveness and cost-efficiency to reveal\footnote{Tests are written/generated to kill (reveal) the mutants. A bug is revealed by a mutation testing approach, if the written tests to kill its mutants also reveal the bug.} them. 
Our results show that \OurTool is very effective in terms of fault revelation, finding on average \changing{84\%} of the faults. 
This implies that \OurTool mutants cover efficiently faulty behaviours caused by real bugs.
More importantly, the approach is noticeably more effective and cost-efficient than a traditional mutation testing technique, namely PiTest~\cite{pitest}, that we use as a baseline in our evaluation. 
Precisely, we consider three different configurations for PiTest that uses different sets of mutation operators (DEFAULT, ALL and RV). 
In fact, test suites that kill all mutants of \OurTool find on average between \changing{5.5\% to 33\%} more faults than those generated to kill all mutants introduced by PiTest. Moreover, even when analysing the same number of mutants, \OurTool induces test suites that find on average \changing{6\% to 16\%} more faults than PiTest. 
These results are promising and endorse the usage of \OurTool over the considered  mutation testing technique, as a test generation and assessment criterion.   

We also study the impact of the condition-seeding-based mutations in the fault detection capability of \OurTool.
We observe that test-suites designed to kill both kinds of \OurTool mutants -- induced by 1) direct CodeBERT predictions and 2) a combination of conditions-seeding with CodeBERT predictions -- find on average over \changing{9\%} more bugs than the ones designed to kill direct CodeBERT prediction mutants only (1).


Overall, our main contributions are: 

\begin{itemize}
    \item We introduce \OurTool, the first mutation testing approach that uses pre-trained language models.
    It leverages the model's code knowledge captured during its pretraining on large code corpora and its ability to capture the program context, to produce ``natural'' mutants.
    
    \item We propose new additive mutations which operate by seeding new conditions in the existing conditional expressions of the target code, then masking and replacing their tokens with the model predictions.
    
     \item We provide empirical evidence that \OurTool mutants can guide testing towards higher fault detection capabilities, outperforming those achieved by SOA techniques (i.e. PiTest), in terms of effectiveness and cost-efficiency. 
     In our empirical study, we validate also the advantage of employing the new additive mutation patterns, w.r.t improving the effectiveness and cost-efficiency in writing test suites with higher fault revelation capability.
\end{itemize}

\section{Background}

\begin{figure*}[htp!]
\centering
\includegraphics[width=\textwidth]{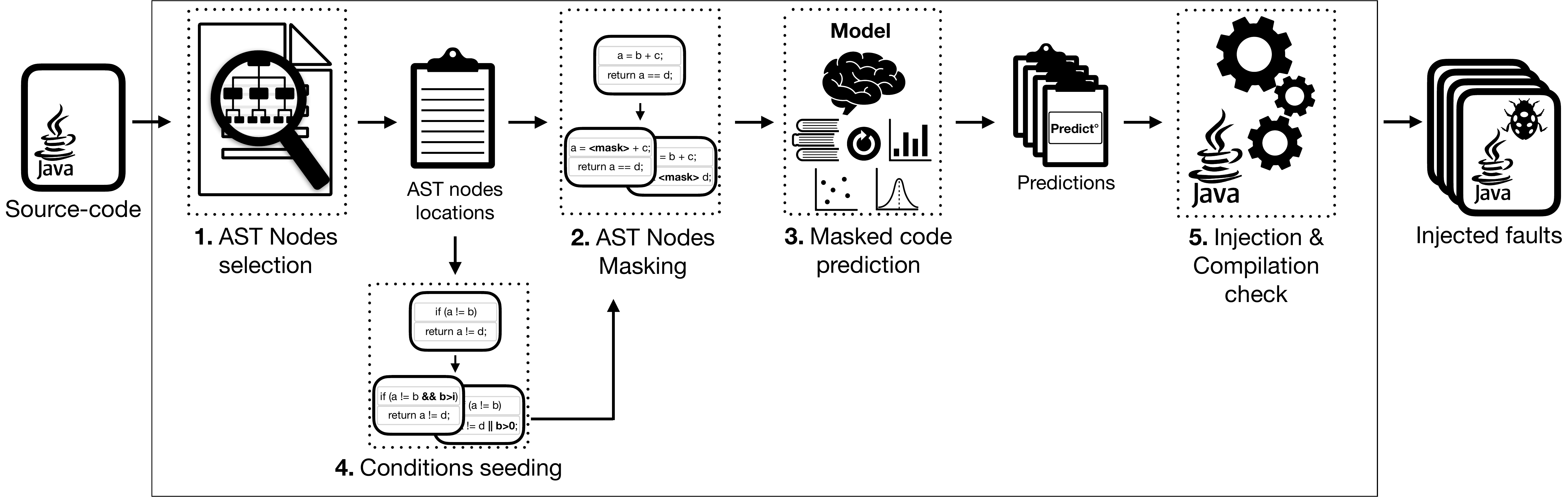}
\caption{{\OurTool} Workflow: (1) it parses the Java code given as input, and extracts the expressions to mutate; (2) it creates simple-replacement mutants by masking the tokens of interest and invoking CodeBERT; (3) it generates the mutants by replacing the masked token with CodeBERT predictions; (4) it generates complex mutants via a) conditions-seeding, b) tokens masking then c) replacing by CodeBERT predictions; and finally, (5) it discards not compiling and syntactically identical mutants.}
\label{fig:muBERT}
\end{figure*}

\subsection{Mutation Testing}
Mutation analysis~\cite{PapadakisK00TH19} is a test adequacy criterion representing test requirements by the mean of mutants, 
which are obtained by performing slight syntactic modifications to the original program.
For instance, an expression like  $\texttt{x > 0}$ can be mutated to $\texttt{x < 0}$ by replacing the relational operator \texttt{>} with \texttt{<}. 
These mutants are then used to assess the effectiveness and thoroughness of a test suite in detecting their corresponding code modification. 

A test case detects a mutant if it is capable of producing distinguishable observable outputs between the mutant and the original program. A mutant is said to be \emph{killed} if it is detected by a test case or a test suite; otherwise, it is called \emph{live} or \emph{survived}. 
Some mutants cannot be killed as they are functionally \emph{equivalent} to the original program. 
The \emph{mutation score} measures the test suite adequacy and is computed as the ratio of killed mutants over the total number of generated mutants. 

\subsection{Generative Language Models}

Advances in deep learning approaches gave birth to new language models for code generation ~\cite{copilot,openAIcodex,codewhisperer,DBLP:conf/emnlp/FengGTDFGS0LJZ20}. 
These models are trained on large corpora counting multiple projects, thereby acquiring a decent knowledge of code, enabling them to predict accurately source code to developers. 
Among these pre-trained models, CodeBERT~\cite{DBLP:conf/emnlp/FengGTDFGS0LJZ20}, a language model that has been recently introduced and made openly accessible for researchers by Microsoft.

CodeBERT is an NL-PL bimodal pre-trained language model (Natural Language Programming Language) that supports multiple applications such as code search, code documentation generation, etc. 
Same as most large pre-trained models, i.e. BERT~\cite{devlin2018bert}, CodeBERT's developing adopts a Multilayer Transformer~\cite{vaswani2017attention} architecture.
It has been trained on a large corpus collected from over 6.4 million projects available on GitHub, counting 6 different programming languages, including Java.
The model was trained in a cross-modal fashion, through bimodal NL-PL data, where the input data is formed by pairs of source code and its related documentation, as well-as unimodal data, including either natural language or programming language sequences per input.
This way, it enables the model to offer both -- PL and NL-PL -- functionalities.
The training targets a hybrid objective function, that is based on replaced token detection.

\OurTool incorporates the Masked Language Modeling (MLM) functionality~\cite{codebertweb} of CodeBERT in its workflow, to generate ``natural'' mutants. 
The CodeBERT MLM pipeline takes as input a code sequence of maximum 512 tokens, including among them one masked as \texttt{<mask>}, whose value will be predicted by the model based on the context captured from the remaining tokens. 
CodeBERT provides by default 5 predictions per token, among which we use the inaccurate and compilable predicted codes as mutants.


\section{Approach}
\label{sec:approach}

We propose \OurTool, a generative language-model-based mutation testing approach, which is described step by step in 
Figure~\ref{fig:muBERT}. 
Given an input source code, \OurTool leverages CodeBERT's knowledge of code and its capability in capturing the program's context to produce ``natural'' mutations, i.e. that are similar to eventual developer mistakes.
To do so, \OurTool proceeds as follows in six steps:

\begin{enumerate}
\item First, it extracts relevant locations (AST~\footnote{AST: Abstract Syntax Tree.} nodes) where to mutate 
\item Second, it masks the identified node-tokens, creating one masked version per selected token.
\item Then, it invokes CodeBERT to predict replacements for these masked tokens.
\item In addition to the mutants produced in Step (3), {\OurTool} also implements some condition-seeding additive mutations that modify more than one token. Precisely, it modifies the conditional expressions in the control flow (typically present in \texttt{if}, \texttt{do}, \texttt{while} and \texttt{return} statements) by extending the original condition with a new one, combined with the logical operator \texttt{\&\&} or \texttt{||}. Then, the new conditional expression is mutated by following the same steps (2) and (3) -- masking and replacing the masked tokens by the CodeBERT predictions. 
\item Finally, the approach discards duplicate predictions or those inducing similar code to the original one, or not compiling, and outputs the remaining ones as mutants, from diverse locations of the target code. 
More precisely, it iterates through the statements in random order and outputs in every iteration one mutant per line, until achieving the desired number of mutants or all mutants are outputted. 
\end{enumerate}

\subsection{AST Nodes Selection}
\OurTool parses the AST of the input source code and selects the lines that are more likely to carry the program's specification implementation, excluding the import statements and the declaration ones, e.g. the statements declaring a class, a method, an attribute, etc. 
This way, the approach focuses the mutation on the business-logic portion of the program and excludes the lines that are probably of lower impact on the program behaviour.
It proceeds then, by selecting from each of these statements, the relevant nodes to mutate, i.e. the operators, the operands, the method calls and variables, etc., and excluding the language-specific ones, like the separators and the flow controls, i.e. semicolons, brackets, \texttt{if}, \texttt{else}, etc.
Table~\ref{tab:mbert_conventional_patterns} summarises the type of targeted AST nodes by {\OurTool}, with corresponding example expressions and induced mutants. 
We refer to these as the conventional mutations provided by {\OurTool}, denoted by {\Toolconv} in our evaluation, previously introduced in the preliminary version of the approach~\cite{DBLP:conf/icst/DegiovanniP22}.

\subsection{Token Masking}
In this step, we mask the selected nodes one by one, producing a masked version from the original source code for each node of interest.
This means that every masked version contains the original code with one missing node, replaced by the placeholder~\texttt{<mask>}. 

This way, \OurTool can generate several mutants in the same program location. For instance, for an assignment expression like $\texttt{res = a + b}$, {\OurTool} will create (potentially 25) mutants from the following masked sequences: 
\begin{itemize}
    \item $\textbf{\texttt{<mask>}} = \texttt{a + b}$
    \item $\texttt{res \textbf{<mask>}= a + b}$
    \item $\texttt{res} = \texttt{\textbf{<mask>} + b}$
    \item $\texttt{res} = \texttt{a \textbf{<mask>} b}$
    \item $\texttt{res} = \texttt{a + \textbf{<mask>}}$
\end{itemize}


\begin{table*}[htp!]
\vspace{0.6em}
\centering
\caption{Example of {\OurTool} conventional mutations, available in the preliminary version of the approach~\cite{DBLP:conf/icst/DegiovanniP22}, denoted by {\Toolconv}.}
\begin{tabular}{lrrr}
\toprule
\textbf{Ast node}  &\textbf{Expression} & \textbf{Masked Expression} & \textbf{Mutant Example}\\
\midrule
literals & \texttt{res + 10} & $\texttt{res + \textbf{<mask>}}$ & $\texttt{res + \textbf{0}}$\\
identifiers & \texttt{res + 10} & $\texttt{\textbf{<mask>} + 10}$ & $\texttt{\textbf{a} + 10}$\\
binary expressions & \texttt{a \&\& b} & $\texttt{a \textbf{<mask>} b}$ & $\texttt{a \textbf{||} b}$\\
unary expressions & \texttt{--a} & $\texttt{\textbf{<mask>}a}$ & $\texttt{\textbf{++}a}$\\
assignments & \texttt{sum += current} & $\texttt{sum \textbf{<mask>}= current}$ & $\texttt{sum \textbf{-}= current}$\\
object fields & \texttt{node.next} & $\texttt{node.\textbf{<mask>}}$ & $\texttt{node.\textbf{prev}}$\\
method calls & \texttt{list.add(node)} & $\texttt{list.\textbf{<mask>}(node)}$ & $\texttt{list.\textbf{push}(node)}$\\
array access & \texttt{arr[index + 1]} & $\texttt{arr[\textbf{<mask>}]}$ & $\texttt{arr[\textbf{index}]}$\\
static type references & \texttt{Math.random() * 10} & $\texttt{\textbf{<mask>}.random() * 10}$ & $\texttt{\textbf{Random}.random() * 10}$\\ 
\bottomrule
\end{tabular}
\label{tab:mbert_conventional_patterns}
\end{table*}

\subsection{CodeBERT-MLM prediction}
\OurTool invokes CodeBERT to predict replacements for the masked nodes.
To do so, it tokenizes every masked version into a tokens vector then crops it to a subset one that fits the maximum size allowed by the model (512) and counts the masked token with the surrounding code-tokens. 
Next, our approach feeds these vectors to CodeBERT MLM to predict the most probable replacements of the masked token.
Our intuition is that the larger the code portion accompanying the mask placeholder, the better CodeBERT would be able to capture the code context, and consequently, the more meaningful its predictions would be. 
This step ends with the generation of five predictions per masked token. 

\subsection{Condition seeding}
\OurTool generates second-order mutants by combining condition seeding with CodeBERT prediction capabilities.
To do so, our approach modifies the conditions in control flow and return statements, including \texttt{if}, \texttt{do}, \texttt{while} and \texttt{return} conditional expressions.
For every one of these statements, it starts by extending the original condition by a new one, separated with the logical operator \texttt{\&\&} or \texttt{$||$}, in both orders (original condition first or the other way around) and with or without negation (\texttt{!}). 

Next, all substitute conditions are put one by one in place in the original code, forming multiple condition-seeded code versions, that we pass as input to Step (2), in which their tokens are masked and then (3) passed each to CodeBERT to predict the best substitute of their corresponding masked tokens.

The seeded conditions are created in two ways:

\subsubsection{Using existing conditions in the same class}
To mutate a given condition -- \texttt{if}, \texttt{do}, \texttt{while} and \texttt{return} conditional expressions --, we collect all other conditions existing in the same class, then combine each one of them with the target condition, using logical operators.

Precisely, let \texttt{$Exp_t$} a conditional expression to mutate and \texttt{$S_E=\{Exp_0, ... , Exp_n\}$} the set of other conditional expressions appearing in the same class, excluding the null-check ones (i.e. \texttt{var == null}).
The alternative replacement conditions generated for \texttt{$Exp_t$} are the combinations of: 
\begin{itemize}
\item \texttt{$Exp_t$ op neg $Exp_i$} and 
\item\texttt{$Exp_i$ op neg $Exp_t$}, 
\end{itemize}
where 
\texttt{op} is a binary logical operator taking the values in \texttt{\{\&\&,||\}}, 
\texttt{neg} is either the negation operator \texttt{!} or nothing and  
\texttt{$Exp_i$} is a condition from {$S_E$}.

\subsubsection{Using existing variables in the same class}
When the target \texttt{if} conditional expression to mutate contains variables (including fields), we create new additional conditions by combining these variables with others of the same type from the same class. Then we combine each one of the newly created conditions with the original one, using logical operators.

Precisely, let \texttt{$Exp_t$} be a conditional expression to mutate  containing a set of variables \texttt{$S_{vt}$}. 
For every variable \texttt{$var_t$} in \texttt{$S_{vt}$}, we load \texttt{$S_v=\{var_0, ... , var_n\}$} the set of other variables appearing in the same class and of the same type \texttt{$T$} as \texttt{$var_t$}, then we generate the following new conditions: 
\begin{itemize}
\item \texttt{$Exp_t$ op ($var_t$ $rel_{op}$ $var_i$)} and 
\item \texttt{($var_t$ $rel_{op}$ $var_i$) op $Exp_t$},
\end{itemize}
where \texttt{op} is a binary logical operator taking the values in \texttt{\{\&\&,||\}}, 
\texttt{$rel_{op}$} is a relational operator applicable on the type \texttt{$T$} and 
\texttt{$var_i$} is a variable from {$S_v$}.



\subsection{Mutant filtering}
In this step, our approach starts by discarding accurate and duplicate predictions; the redundant predictions and the ones that are exactly the same as the original code.
Then, it iterates through the statements and selects in every iteration one compilable prediction by line, while discarding not compilable ones.
Once all first-order mutants are selected (issued by one single token replacement), our approach proceeds by selecting second-order ones (issued by the combination of condition seeding and one token replacement) in the same iterative manner.
\OurTool continues iterating until achieving the desired number of mutants or all mutants are outputted.

\section{Research Questions}
We start our analysis by investigating the advantage brought by the additive mutations (a.k.a. conditions seeding ones) w.r.t. the fault detection capabilities of test suites designed to kill {\OurTool}'s mutants. Thus, we ask:

\begin{description}
\item[RQ1] \emph{(\OurTool Additive mutations)} What is the added value of the additive mutations on the fault detection capabilities of test suites designed to kill {\OurTool}'s mutants?
\end{description}

To answer this question, we generate two sets of mutants using {\OurTool}: 1) the first set using all possible mutations that we denote as {\OurTool} and 2) a second one using only the conventional {\OurTool}' mutations -- part of our preliminary implementation~\cite{DBLP:conf/icst/DegiovanniP22}, excluding the additive ones  -- that we denote as {\Toolconv}. 
Then we evaluate the fault detection ability of test suites selected to kill the mutants from each set. 


The answer of this question provides evidence that the additive mutations increase the fault detection capability of \OurTool. 
Yet, to assess its general performance we compare it to state-of-the-art (SOA) mutation testing, particularly PiTest~\cite{pitest}, and thus, we ask:

\begin{description}
\item[RQ2] \emph{(Fault detection)} How does {\OurTool} compare with state-of-the-art mutation testing, in terms of fault detection?
\end{description}

To answer this question we generate mutants using the latest version of PiTest~\cite{pitest}, on the same target projects as for RQ1. As we are interested in comparing the approaches and not the implementations of the tools, we exclude the subjects on which PiTest did not run correctly or did not generate any mutant. This way we ensure having a fair base of comparison by counting exactly the same study subjects for both approaches (further details are given in Section~\ref{sec:setup}). Then, we compare the fault detection capability of test suites selected to kill the same number of mutants produced by each approach.

Finally, we qualitatively analyse some of the mutants generated with {\OurTool} and ask:

\begin{description}
\item[RQ3] \emph{(Qualitative analysis)} Does {\OurTool} generate different mutants than traditional mutation testing operators?
\end{description}

To answer this question, we showcase the mutants generated by {\OurTool} that help in detecting faults not found by PiTest. 
Additionally, we discuss the program-context-capturing importance in \OurTool's functioning, by rerunning it with a reduced size of the masked codes passed to the model, and comparing examples of yielded mutants with those obtained in our original setup. 


\section{Experimental Setup}
\label{sec:setup}

\subsection{Dataset \& Benchmark}

To evaluate \OurTool's fault detection, we use real bugs from a popular dataset in the software engineering research area -- Defects4J~\cite{defects4j} v2.0.0. 
In this benchmark, every subject bug is provided with a buggy version of the source code, its corresponding fixed version, and equipped with a test suite that passes on the fixed version and fails with at least one test on the buggy one. 
The dataset includes over 800 bugs from which, we exclude the ones presenting issues, i.e. with wrong revision ids, not compiling or with execution issues, or having failing tests on the fixed version, at the reporting time. 
Next, we run \OurTool and PiTest on the corresponding classes impacted by the bug from the fixed versions of the remaining bugs and exclude the ones where no tool generated any mutant, ending up with \changing{689} bugs covered by \OurTool and \changing{457} covered by PitTest.
As we're interested in comparing the approaches and not the tools' implementations, and to exclude eventual threats related to the environment (i.e. supported java and juint versions by each technique, etc.) or the limitations and shortages of the dataset, we establish every comparison study on a dataset counting only bugs covered by all considered approaches: \changing{689} bugs to answer RQ1 and \changing{457} to answer RQ2 and RQ3.  

\subsection{Experimental Procedure}
\label{subsec:exp_proc}


To assess the complementary and added value in terms of fault revelation of the condition-seeding-based mutations (answer to RQ1), we 
run our approach with and without those additional mutations -- that we name respectively \OurTool and \Toolconv --, and thus, generating all possible mutants on our dataset programs' fixed versions. 
Next, we compare the average effectiveness of the test suites generated to kill the mutants of each set; induced by \OurTool and \Toolconv.   


Once the added value of the proposed condition-seeding-based mutations is validated, we compare its performance to S.O.A. mutation testing (answer to RQ2 and RQ3). 
We use PiTest~\cite{pitest}, a stable and mature Java mutation testing tool, because it has been more effective at finding faults than other tools \cite{KintisPPVMT18} and it is among the most commonly used by researchers and practitioners \cite{PapadakisK00TH19, SanchezDMS22}, as of today. 
The tool proposes different configurations to adapt the produced mutations and their general cost to the target users, by excluding or including mutators.
Among these configurations we used the three following:
\begin{itemize}
\item Pit-all (\texttt{ALL}) which counts all available mutation operators available in the current version\footnote{Version 1.9.4 available in PitTest's~\cite{pitweb} GitHub repository (branch=master, repo=\url{https://github.com/hcoles/pitest.git}, rev-id=17e1eecf)}.
\item Pit-default (\texttt{DEFAULTS}) whose mutators are selected to form a stable and cost-efficient subset of operators by producing less but more relevant mutants.
\item Pit-rv-all (\texttt{ALL}) which is a version\footnote{Version 1.7.4 available in PitTest's~\cite{pitweb} GitHub repository (branch=master, repo=\url{https://github.com/hcoles/pitest.git}, rev-id=2ec1178a)} that includes the mutators of Pit-all and extra experimental~\cite{pit-rv-web} ones that are made available for research studies.
\end{itemize}

To compare the different approaches, we evaluate their effectiveness and cost-efficiency in achieving one of the main purposes of mutation testing, i.e., to guide the testing towards higher fault detection capabilities.
For this reason, we simulate a mutation testing use-case scenario, where a developer/tester selects mutants and writes tests to kill them~\cite{ChekamPBTS20, KurtzAODKG16}.

We run every approach on the fixed versions and test suites provided by Defects4J, then collect the mutants and their test execution results; whether the mutant is killed (breaks at least one test of the test suite) and if yes by which tests.
Next, we suppose that the not killed mutants are equivalent or irrelevant, explaining why no tests have been written to kill them by the developers. 
Then, we simulate the scenario of a developer testing the fixed version, in a state where 1) it did not have any test 2) thus all mutants did not have killing tests and 3) the developer had no knowledge of which mutants are equivalent or not.
This way, we can reproduce the developer flow of
\begin{enumerate}
\item selecting and analysing one mutant,
\item to either (a) discard it from the mutant set if it is equivalent (not killed in the actual test suite) or (b) write a test to kill it (by selecting one of the actual killing tests of the mutant),
\item then discarding all killed mutants by that test and 
\item iterating similarly over the remaining mutants until all of them are analysed.
\end{enumerate}
We say that a bug is found by a mutation testing technique if the resulting test suite -- formed by the written (selected) tests by the developer -- contains at least one test that reveals it; a test that breaks when executed on the buggy version.

We express the testing cost in terms of mutants analysed, and hence, we consider the effort required to find a bug as the number of mutants analysed until the first bug-revealing test is written.
To set a common basis of comparison between the approaches, accounting for the different number of generated mutants, we run the simulations until the same maximum effort is reached (maximum number of mutants to analyse), which we set to the least cost required to kill all the mutants by one of the compared approaches.
During our evaluation study, we use the same mutation selection strategy for all compared approaches, iterating through the lines in random order and selecting 1 arbitrary mutant per line per iteration. 
To reduce the process randomness impact on our results (in the selection of mutants and tests), we run every simulation 100 times, then average their results for every target-bug and considered approach. 
Finally, we aggregate these averages computed on all target bugs and normalise them as global percentages of achieved fault detection by spent effort, in terms of mutants analysed.


Finally, to answer RQ3, we select example mutants that enabled \OurTool to find bugs exclusively (not found by any of PiTest versions), from the results of RQ2. 
Then we discuss the added value of \OurTool mutations through the analysis of the mutants' behavioural difference from the fixed version and similarity with the buggy one.

\subsection{Implementation}
We implemented \OurTool's approach as described in Section~\ref{sec:approach}: we have used Spoon~\cite{pawlak:hal-01169705} and Jdt~\cite{eclipse2013eclipse} libraries to parse and extract the business logic related AST nodes and apply condition-seeding mutators. 
To predict the masked tokens we have used the implementation proposed by CodeBERT-nt~\cite{khanfir2022codebert, codebertntweb}, using CodeBERT Masked Language Modeling (MLM) functionality~\cite{DBLP:conf/emnlp/FengGTDFGS0LJZ20,codebertweb}.

We provide the implementation of our approach and the reproduction package of its evaluation at \url{https://github.com/Ahmedfir/mBERTa}.

\section{Experimental Results}
\subsection{RQ1: \OurTool Additive mutations}

         

\begin{figure}[t]
\centering 
     \begin{subfigure}[b] {0.48\textwidth}
         \centering
         \adjincludegraphics[width=\textwidth, trim={{.04\width} {.017\width} {0.08\width} {.08\width}} ,clip]{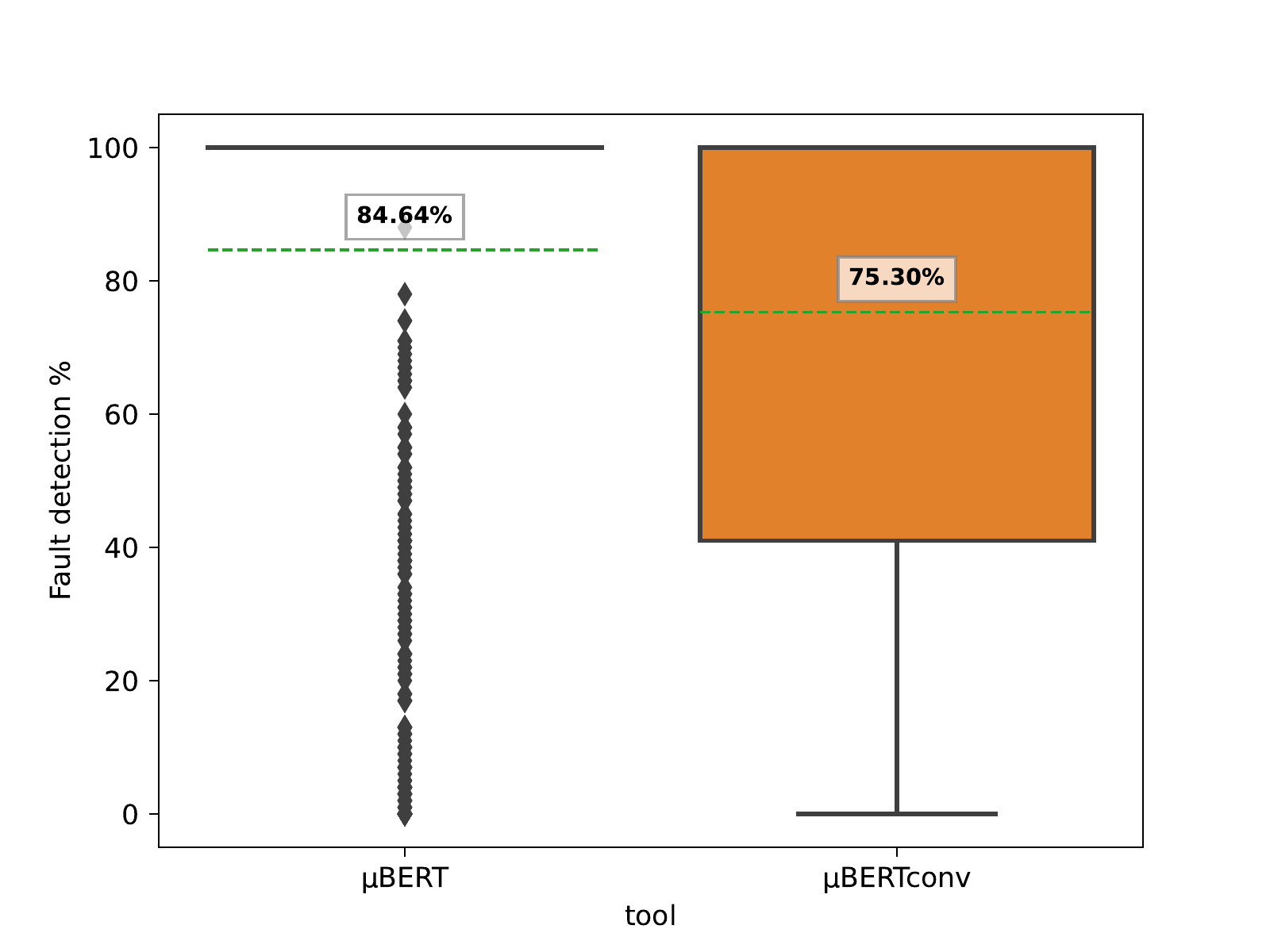}
         \caption{Effectiveness: mean fault-detection per subject.}
         \label{fig:box-plot-mbert-all}
     \end{subfigure}
     \hfill
     \vspace{1em}
     \begin{subfigure}[b] {0.48\textwidth}
         \centering
         \adjincludegraphics[width=\textwidth, trim={{.04\width} {.017\width} {0.08\width} {.08\width}} ,clip]{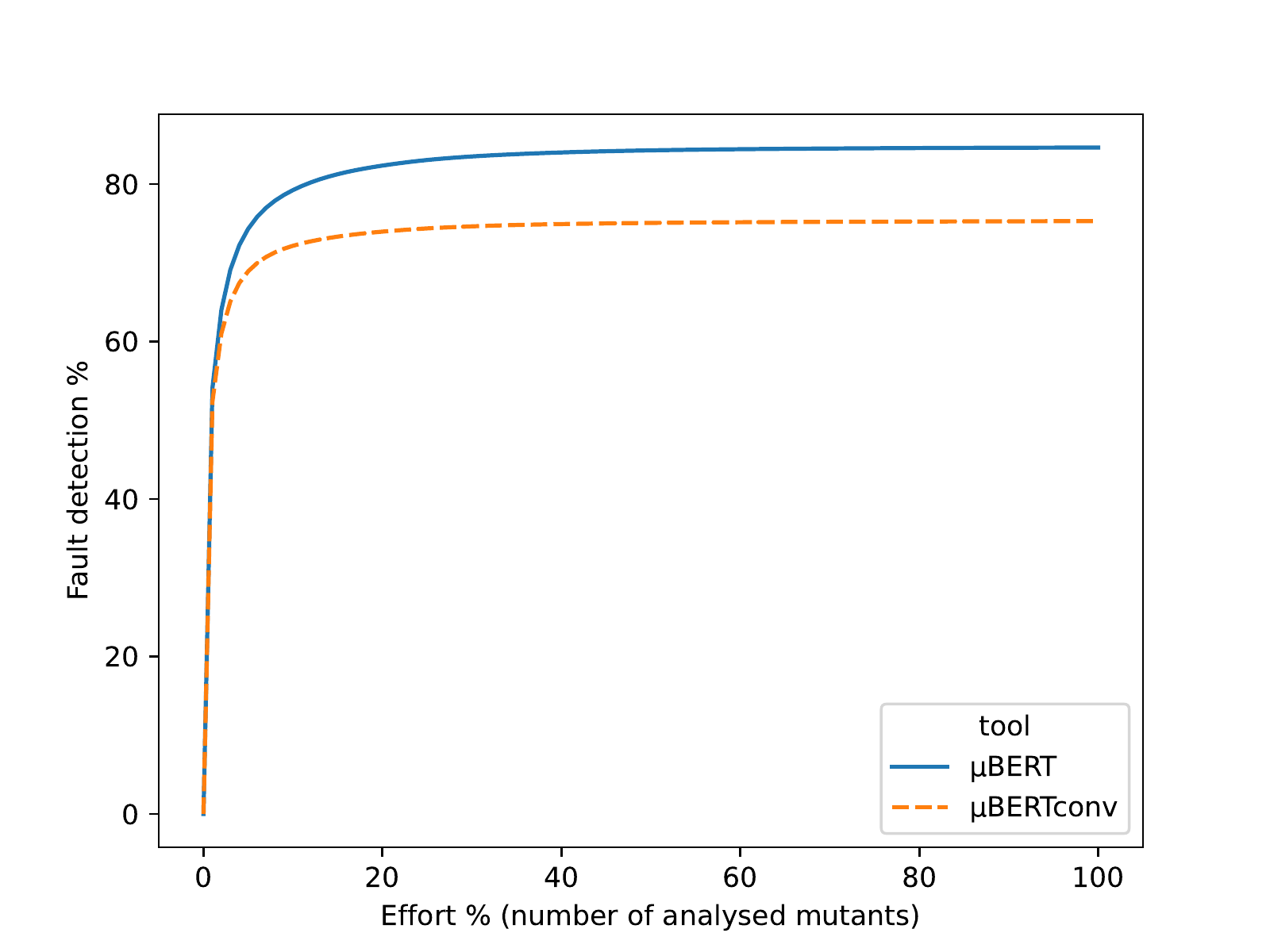}
         \caption{Cost-efficiency: fault detection by the number of mutants analysed.}
         \label{fig:line-plot-mbert-all}
     \end{subfigure}
     \hfill     
     \caption{Fault-detection performance improvement when using additive patterns. Comparison between \OurTool and \Toolconv, w.r.t. the fault-detection of test suites written to kill \textbf{all} generated mutants.}
    \label{fig:fd-bp-additive-patterns}
\end{figure}

To answer this question we compare the fault detection effectiveness of test suites written to kill mutants generated by \OurTool with and without additive mutations, noted respectively \OurTool and \Toolconv. 
Figure~\ref{fig:fd-bp-additive-patterns} depicts the fault detection improvement when extending \OurTool mutations by the additive ones. 
In fact, \OurTool fault detection increased on average by over \changing{9\%} compared to the one achieved by \Toolconv, achieving \changing{84.64\%} on average. 
We can also see that besides outliers, the majority of bugs are found in \changing{100\%} of the times. 
Moreover, when examining the bugs separately, we find that \OurTool finds \changing{20} more bugs than \Toolconv (with fault detection $> 0\%$), and \changing{70} more when considering bugs found with fault detection percentages above 90\%. 
This confirms that the additive patterns induce relevant mutants ensuring the detection of some bugs always or in most of the cases, as well as representing better new types of faults, which were not detectable otherwise.

To check the significance of the fault detection advantage brought by the additive patterns, we performed a statistical test (Wilcoxon paired test) on the data of Figure~\ref{fig:box-plot-mbert-all} to validate the hypothesis "the fault detection yielded by \OurTool is \textbf{greater} than the one by \Toolconv". The very small obtained p-values of  \changing{5.92e-21} ($\ll$ \changing{0.05}) showed that the differences are significant, indicating the low probability of this fault detection amelioration to be happening by chance. 
The difference size confirms also the same advantage, with $\hat{\text{A}}_{12}$ values of \changing{0.5827} ($>$ \changing{0.5}), indicating that \OurTool induces test-suites with higher fault detection capability in the majority of the cases.

\begin{figure}[t]
\centering 
     \begin{subfigure}[b] {0.48\textwidth}
         \centering
         \adjincludegraphics[width=\textwidth, trim={{.04\width} {.017\width} {0.08\width} {.08\width}} ,clip]{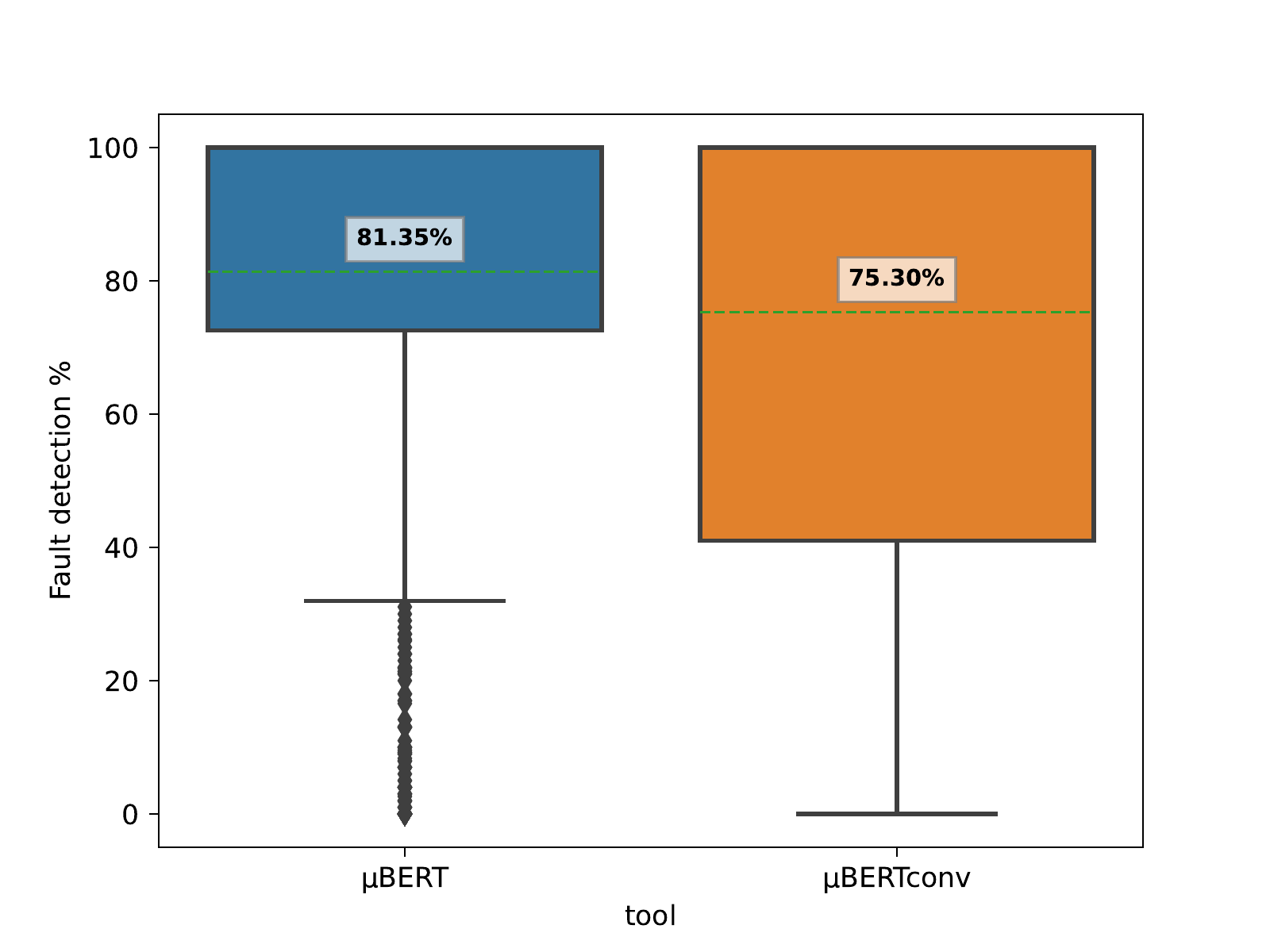}
         \caption{Effectiveness: mean fault-detection per subject.}
         \label{fig:box-plot-mbert-fix}
     \end{subfigure}
     \hfill
     \vspace{1em}
     \begin{subfigure}[b] {0.48\textwidth}
         \centering
         \adjincludegraphics[width=\textwidth, trim={{.04\width} {.017\width} {0.08\width} {.08\width}} ,clip]{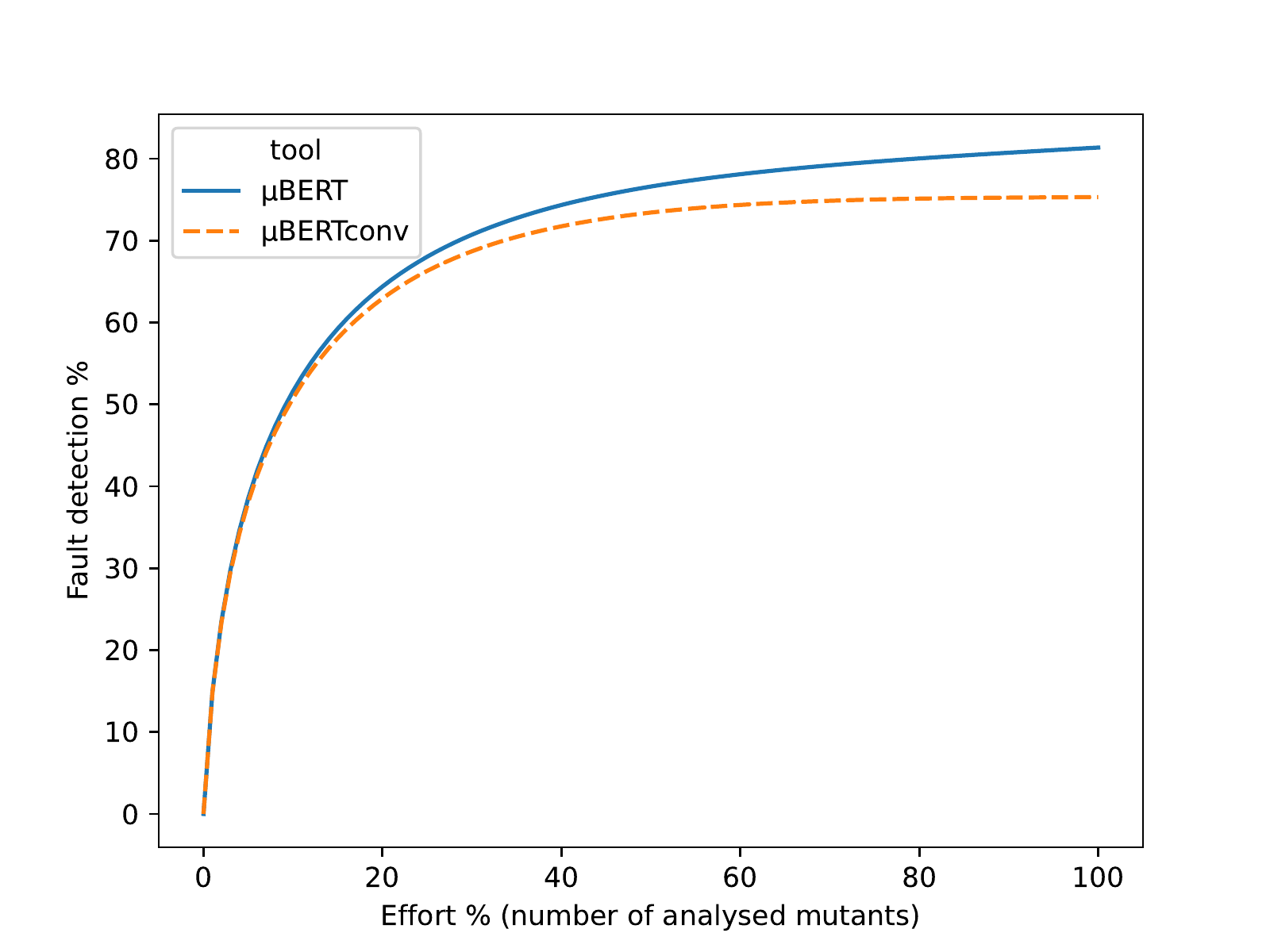}
         \caption{Cost-efficiency: fault detection by the number of mutants analysed.}
         \label{fig:line-plot-mbert-fix}
     \end{subfigure}
     \hfill     
     \caption{Fault-detection comparison between \OurTool and \Toolconv, with the \textbf{same effort}: \small{where the maximum effort is limited to the minimum effort required to analyse all mutants of any of them, which is \Toolconv in most of the cases.} }
    \label{fig:fd-mbert}
\end{figure}

Next, we compare the fault detection performance of \OurTool and \Toolconv when analysing the same number of mutants, and illustrate in Figure~\ref{fig:fd-mbert} their average fault detection effectiveness and cost-efficiency in terms of analysed mutants. 
The box-plots of the Subfigure~\ref{fig:box-plot-mbert-fix} show that even when spending the same effort as \Toolconv, \OurTool keeps a similar advantage of on average \changing{6.05\%} higher fault detection, achieving a maximum of \changing{81.35\%}.
From the line-plots of the Subfigure~\ref{fig:line-plot-mbert-fix}, we can see that both approaches achieve a comparable fault detection \changing{($\approx70\%$)} at \changing{($\leq\approx40\%$)} of the maximum costs. 
At higher costs, \Toolconv's curve increases slowly until achieving a plateau at \changing{$\approx60\%$} of the effort, whereas \OurTool's curve keeps increasing towards higher fault detection ratios even when achieving the \changing{$\approx100\%$} of the fixed maximum effort.

To validate these findings we re-conducted the same statistical tests on the data of Subfigure~\ref{fig:box-plot-mbert-fix} and found that \OurTool outperforms significantly \Toolconv with negligible p-values of \changing{1.15e-19} and $\hat{\text{A}}_{12}$ values of \changing{0.5711}.

\subsection{RQ2: Fault Detection comparison with PiTest}

\begin{figure}[t]
\centering 
     \begin{subfigure}[b] {0.48\textwidth}
         \centering
         \adjincludegraphics[width=\textwidth, trim={{.04\width} {.017\width} {0.08\width} {.08\width}} ,clip]{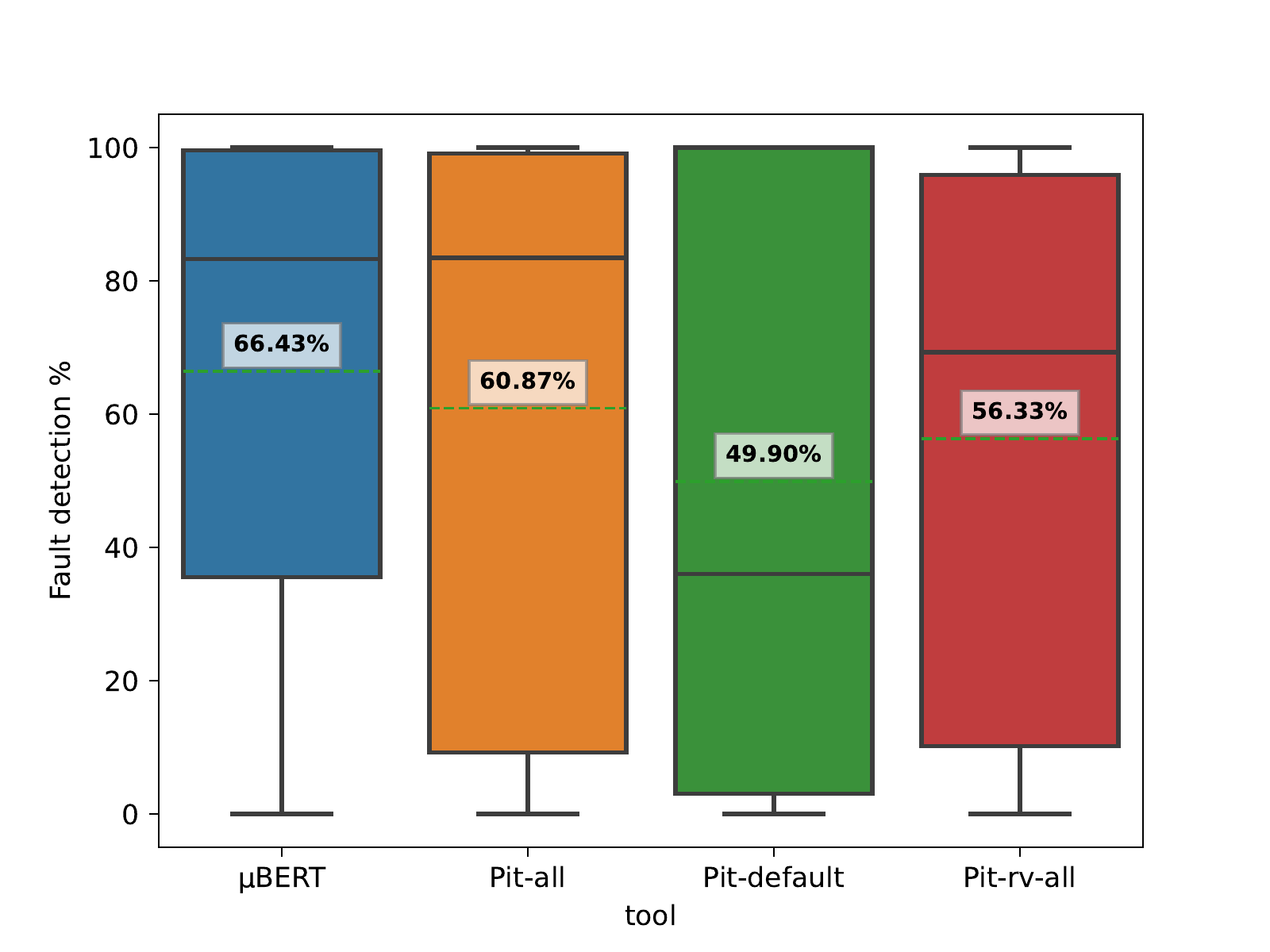}
         \caption{Effectiveness: mean fault-detection per subject.}
         \label{fig:box-plot-pit-fix}
     \end{subfigure}
     \hfill
     \vspace{1em}
     \begin{subfigure}[b] {0.48\textwidth}
         \centering
         \adjincludegraphics[width=\textwidth, trim={{.04\width} {.017\width} {0.08\width} {.08\width}} ,clip]{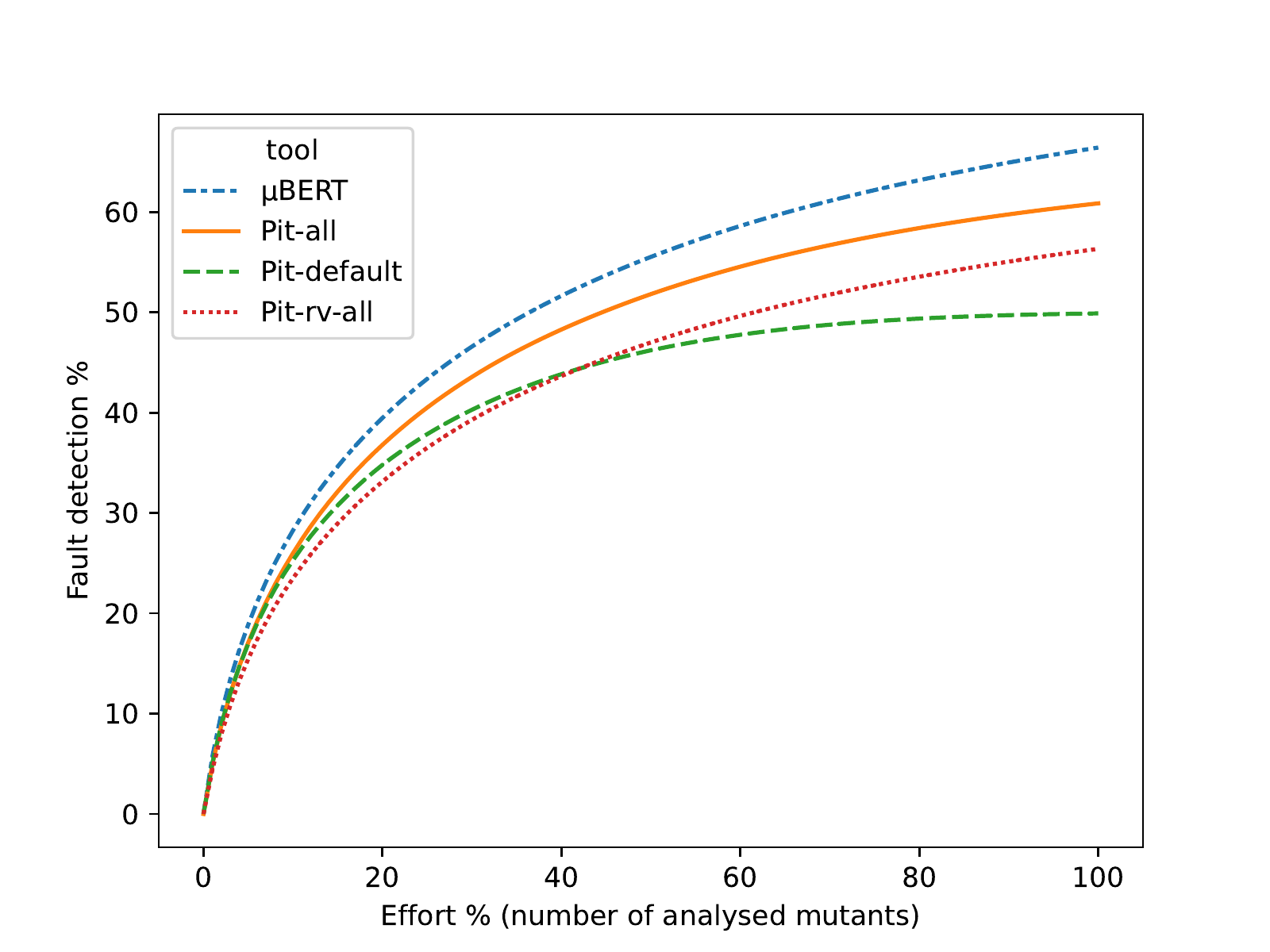}
         \caption{Cost-efficiency: fault detection by the number of mutants analysed.}
         \label{fig:line-plot-pit-fix}
     \end{subfigure}
     \hfill     
     \caption{Fault-detection comparison between \OurTool and PiTest, with the \textbf{same effort}: \small{where the maximum effort is limited to the minimum effort required to analyse all mutants of any of them, which is Pit-default in most of the cases.} }
    \label{fig:fd-pit}
\end{figure}

To answer this research question we reduce our dataset to the bugs covered by \OurTool and the 3 considered versions of PitTest approaches: 
"Pit-default" which contains the default mutation operators of PiTest, "Pit-all" containing all PiTest operators including the default ones and "Pit-rv-all" which contains \textit{experimental} operators~\cite{pit-rv-web} in addition to the "Pit-all" ones.
Then, we perform the same study as in RQ1, where we compare the considered approaches' effectiveness and cost-efficiency based on the fault detection capability of test suites written to kill their generated mutants.
To have a fair base of comparison, we compare the approaches under the same effort in analysing mutants, which is equal to the least average effort required to kill all mutants of one of the approaches (which is the one of Pit-default in the majority of the cases).
As we are interested in comparing the mutation testing approaches and not mutant selection strategies, we run the simulation with the same one-mutant-per-line random sampling of mutants for all techniques (see Subsection~\ref{subsec:exp_proc}).

Figure~\ref{fig:line-plot-pit-fix} shows that with small effort \changing{($\leq\approx5\%$)} all approaches yield comparable fault detection scores \changing{($\approx10\%$)}. However, the difference becomes more noticeable when spending more effort, with \OurTool outperforming all versions of PiTest; achieving on average \changing{16.53\%} higher fault detection scores than Pit-default, \changing{10.10\%} higher than Pit-rv-all and \changing{5.56\%} higher than Pit-all (see Figure~\ref{fig:box-plot-pit-fix}).

\begin{table}[!ht]
    \vspace{0.7em}
    \caption{Paired (per subject bug) statistical tests of the average fault detection of test suites written to kill the \textbf{same number of mutants} generated by each approach (data of Figure~\ref{fig:box-plot-pit-fix}).}
    \begin{subtable}[b]{0.48\textwidth}
        \centering
\caption{Wilcoxon paired test p-values computed on every dataset subject, comparing each approach (A1) from the first column to the other approaches (A2). p-values smaller than 0.05 indicate that (A1) yields an average fault detection significantly higher than that of (A2).}
\label{wilcoxon_comp-mbert_add_rand_1_per_fl__pit_rand_1_per_fl__pit_default_rand_1_per_fl__pit_rv_rand_1_per_fl_plot}
\begin{tabular}{lrll}
\toprule
   p-values &  Pit-rv-all & Pit-default &  Pit-all \\
\midrule
  $\mu$BERT &    7.78e-11 &    1.18e-12 & 3.32e-02 \\
    Pit-all &    1.54e-22 &    8.87e-06 &       -- \\
Pit-default &    9.55e-01 &          -- &       -- \\
\bottomrule
\end{tabular}

    \vspace{1em}
    \end{subtable}
   
    \begin{subtable}[b]{0.48\textwidth}
        \centering
\caption{Vargha and Delaney $\hat{\text{A}}_{12}$ values computed on every dataset subject, comparing each approach (A1) from the first column to the other approaches (A2). $\hat{\text{A}}_{12}$ values higher than 0.5 indicate that (A1) yields an average fault detection higher than that of (A2) in the majority of the cases.}
\label{a12_comp-mbert_add_rand_1_per_fl__pit_rand_1_per_fl__pit_default_rand_1_per_fl__pit_rv_rand_1_per_fl_plot}
\begin{tabular}{lrll}
\toprule
$\hat{\text{A}}_{12}$ &  Pit-rv-all & Pit-default & Pit-all \\
\midrule
            $\mu$BERT &      0.6488 &      0.5514 &  0.5066 \\
              Pit-all &      0.7210 &      0.4956 &      -- \\
          Pit-default &      0.5449 &          -- &      -- \\
\bottomrule
\end{tabular}

    \end{subtable}
    
    \label{tab:comp_a12_pval_rq2}
\end{table}

To validate these results, we performed the same statistical tests as in RQ1, checking the hypothesis that "\OurTool yields better fault detection capabilities than the other approaches". 
We illustrate in the first row of Tables~\ref{wilcoxon_comp-mbert_add_rand_1_per_fl__pit_rand_1_per_fl__pit_default_rand_1_per_fl__pit_rv_rand_1_per_fl_plot} and~\ref{a12_comp-mbert_add_rand_1_per_fl__pit_rand_1_per_fl__pit_default_rand_1_per_fl__pit_rv_rand_1_per_fl_plot} the corresponding computed Wilcoxon paired test p-values and Vargha and Delaney $\hat{\text{A}}_{12}$ values.
Our results show that \OurTool has a significant advantage over the considered SOA approaches with p-values under \changing{0.05}. 
Additionally, \OurTool scores $\hat{\text{A}}_{12}$ values above \changing{0.5} which confirms that guiding the testing by \OurTool mutants instead of those generated by SOA techniques yields comparable or higher fault detection ratios, in the majority of the cases.
Indeed, the $\hat{\text{A}}_{12}$ difference between Pit-all and \OurTool is small (\changing{0.5066}), indicating that both approaches perform similarly or better on some studied subjects and worst on others.

\begin{figure}[t]
\centering 
     \begin{subfigure}[b] {0.48\textwidth}
         \centering
         \adjincludegraphics[width=\textwidth, trim={{.04\width} {.017\width} {0.08\width} {.08\width}} ,clip]{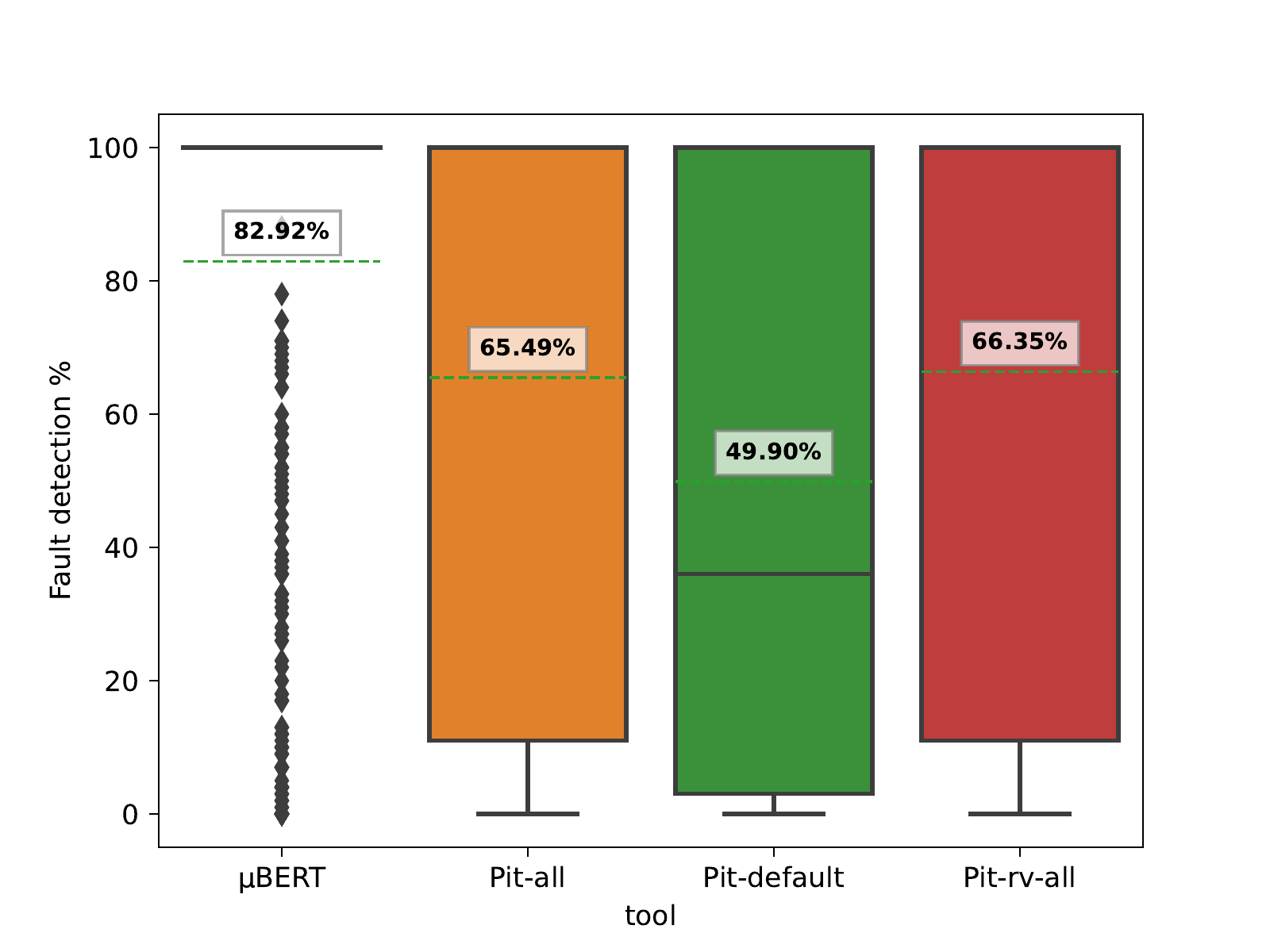}
         \caption{Effectiveness: mean fault-detection per subject.}
         \label{fig:box-plot-pit-all}
     \end{subfigure}
     \hfill
     \vspace{1em}
     \begin{subfigure}[b] {0.48\textwidth}
         \centering
         \adjincludegraphics[width=\textwidth, trim={{.04\width} {.017\width} {0.08\width} {.08\width}} ,clip]{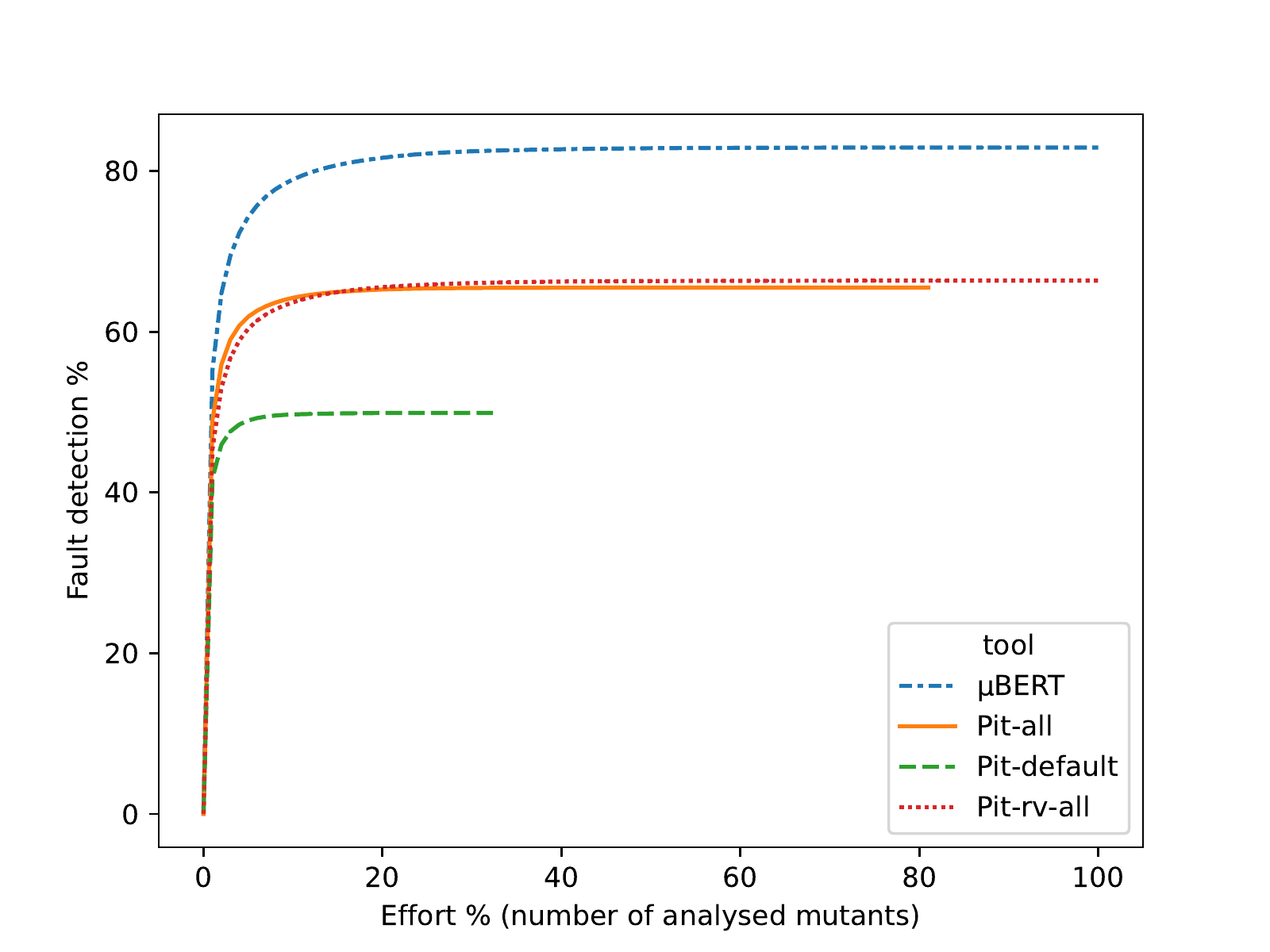}
         \caption{Cost-efficiency: fault detection by the number of mutants analysed.}
         \label{fig:line-plot-pit-all}
     \end{subfigure}
     \hfill     
     \caption{Comparison between \OurTool and PiTest, relative to the fault-detection of test suites written to kill \textbf{all} generated mutants.}
    \label{fig:fd-pit-all}
\end{figure}

We notice also from the sub-figure~\ref{fig:line-plot-pit-fix} that Pit-default achieves a plateau at around \changing{60\%} of the effort while the other tools keep increasing, showing that they are able to achieve higher fault detection capabilities, at a higher cost.
This is very noticeable when we compare the sub-figures (a) and (b) of Figure~\ref{fig:fd-pit} with the figure~\ref{fig:fd-bp-additive-patterns}, where the average fault detection of \OurTool is way lower than what it achieves in RQ1 -- around \changing{66\%} instead of \changing{84\%}.
This is a direct consequence of the fact that Pit default produces fewer mutants than the other approaches, limiting considerably the maximum effort of the mutation campaigns and thus the fault detection ratios, in the majority of the cases.
Indeed, as illustrated in Figure~\ref{fig:fd-pit-all}, all approaches score higher fault detection percentages when spending more effort, achieving on average \changing{$\approx$65\%} for Pit-all, \changing{$\approx$66\%} for Pit-rv-all and \changing{$\approx$83\%} for \OurTool.
We explain the small decrease of \changing{1.72\%} in the mean fault detection achieved by \OurTool in comparison with RQ1 (\changing{82,92\%} in RQ2 instead of \changing{84.64\%} in RQ1) by the difference in the considered dataset for each RQ.

\begin{table}[!ht]
    \vspace{0.7em}
    \caption{Paired (per subject bug) statistical tests of the average fault detection of test suites written to kill \textbf{all the mutants} generated by each approach (data of Figure~\ref{fig:box-plot-pit-all}).}
    \begin{subtable}[b]{0.48\textwidth}
        \centering
\caption{Wilcoxon paired test p-values computed on every dataset subject, comparing each approach (A1) from the first column to the other approaches (A2). p-values smaller than 0.05 indicate that (A1) yields an average fault detection significantly higher than that of (A2).}
\label{wilcoxon_all_comp-mbert_add_rand_1_per_fl__pit_rand_1_per_fl__pit_default_rand_1_per_fl__pit_rv_rand_1_per_fl_plot}
\begin{tabular}{lrll}
\toprule
   p-values &  Pit-rv-all & Pit-default &  Pit-all \\
\midrule
  $\mu$BERT &    2.49e-13 &    2.14e-33 & 1.47e-14 \\
    Pit-all &    4.71e-01 &    2.76e-23 &       -- \\
Pit-default &    1.00e+00 &          -- &       -- \\
\bottomrule
\end{tabular}

        \vspace{1em}
    \end{subtable}
   
    \begin{subtable}[b]{0.48\textwidth}
        \centering
\caption{Vargha and Delaney $\hat{\text{A}}_{12}$ values computed on every dataset subject, comparing each approach (A1) from the first column to the other approaches (A2). $\hat{\text{A}}_{12}$ values higher than 0.5 indicate that (A1) yields an average fault detection higher than that of (A2) in the majority of the cases.}
\label{a12_all_comp-mbert_add_rand_1_per_fl__pit_rand_1_per_fl__pit_default_rand_1_per_fl__pit_rv_rand_1_per_fl_plot}
\begin{tabular}{lrll}
\toprule
$\hat{\text{A}}_{12}$ &  Pit-rv-all & Pit-default & Pit-all \\
\midrule
            $\mu$BERT &      0.6028 &      0.7123 &  0.6061 \\
              Pit-all &      0.5077 &      0.6400 &      -- \\
          Pit-default &      0.3676 &          -- &      -- \\
\bottomrule
\end{tabular}

    \end{subtable}
    
    \label{tab:all_comp_a12_pval_rq2}
\end{table}

In Table~\ref{tab:all_comp_a12_pval_rq2}, we illustrate the $\hat{\text{A}}_{12}$ and p-values computed on data of the boxplots in Sub-figure~\ref{fig:box-plot-pit-all}. The results confirm that \OurTool outperforms significantly SOA mutation testing w.r.t the fault detection capability of test suites written to all kill mutants generated by each approach. 


\begin{figure}[t]
\centering 
     \begin{subfigure}[b] {0.48\textwidth}
         \centering
        \adjincludegraphics[height=6cm,trim={{.18\width} {.22\width} {.11\width} {.13\width}},clip]
         {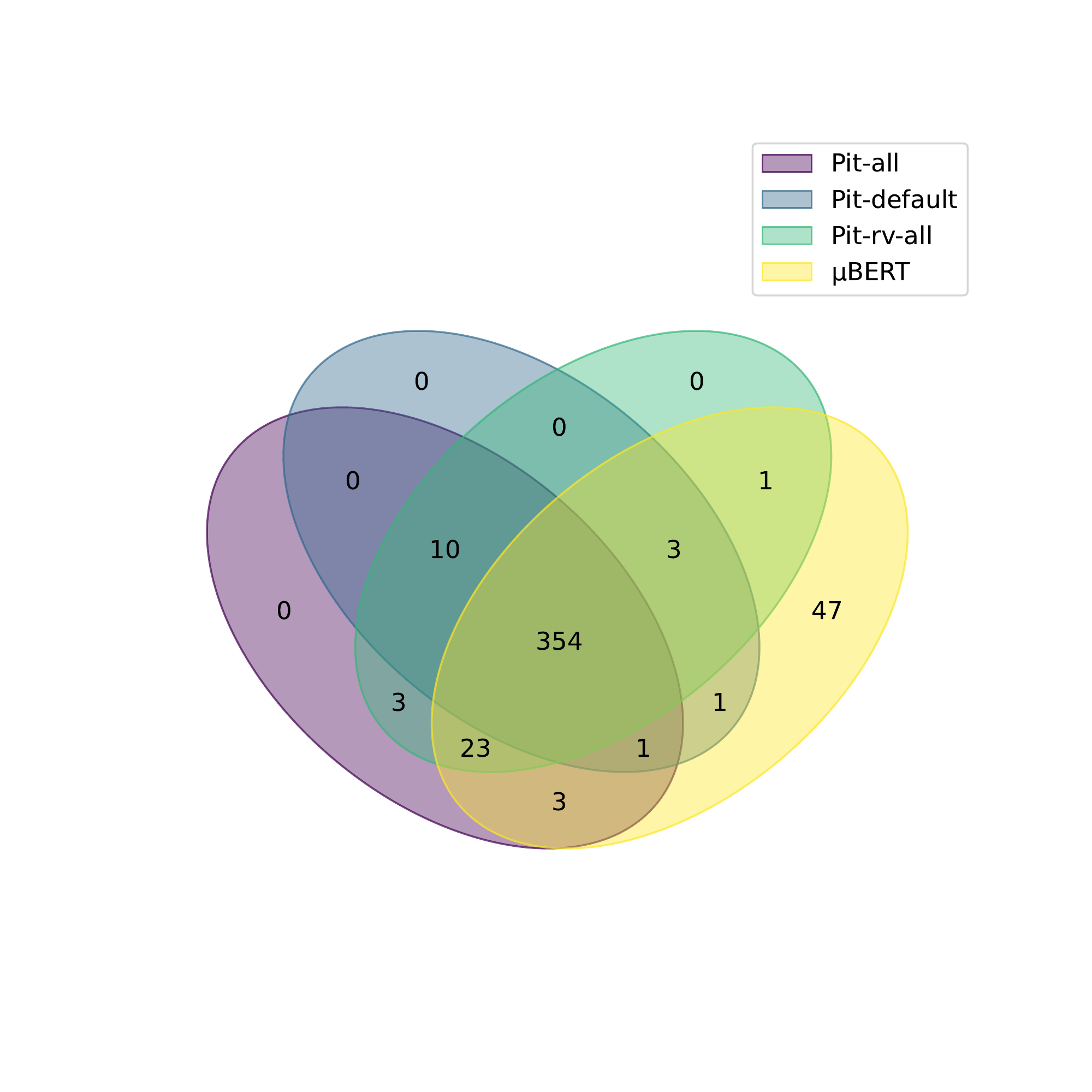} 
         \caption{Faults discovered at least once per 100 runs \newline         (Fault detection $>0\%$).}
        \label{fig:venn0-plot-pit}
     \end{subfigure}
     \hfill
     \begin{subfigure}[b] {0.48\textwidth}
         \centering
         \adjincludegraphics[height=6cm,trim={{.18\width} {.22\width} {.11\width} {.13\width}},clip]
         {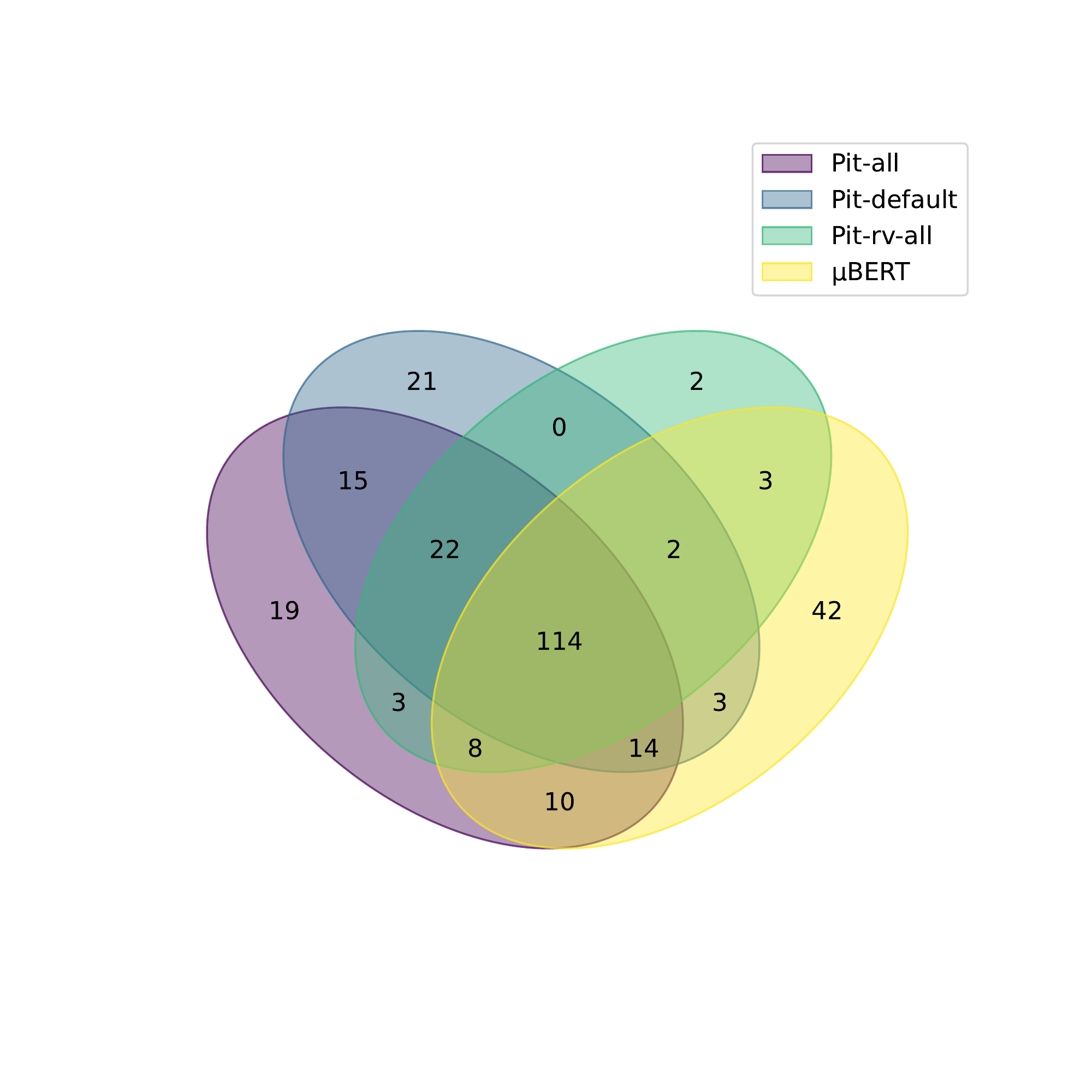} 
         \caption{Faults discovered in over 90\% of the runs \newline 
         (Fault detection$\geq90\%$).}
        \label{fig:venn90-plot-pit}
     \end{subfigure}
     \hfill     
     \caption{Number of faults discovered by test-suites written to kill mutants generated by \OurTool and PiTest versions when analysing the \textbf{same number of mutants} (same effort).}
     \label{fig:venn-plot-pit}
\end{figure}

Next, we turned our interest to the set of particular bugs that every approach can and cannot reveal when spending the same effort.
Hence, we map each bug with its revealing tool, from the simulation results of Figure~\ref{fig:box-plot-pit-fix} and illustrate their corresponding Venn diagrams in Figure~\ref{fig:venn-plot-pit}.

From the  disjoint sets in Sub-figure~\ref{fig:venn0-plot-pit}, we notice a clear advantage in using \OurTool over the considered SOA baselines, as it finds most of the bugs they find in addition to finding exclusively \changing{47} bugs when spending the same effort.
More precisely, \OurTool finds \changing{52}, \changing{77} and \changing{52} more bugs than  Pit-all, Pit-default and Pit-rv-all, respectively, whereas they find each \changing{13}, \changing{10} and \changing{13} bugs that \OurTool missed.

This endorses the fact that \OurTool introduces mutants that represent more real bugs than SOA mutation techniques, and encourages the investigation of the eventual complementary between the approaches.
This observation is more noticeable when considering the overlapping between bugs found by each approach in at least 90\% of the simulations (Sub-figure~\ref{fig:venn90-plot-pit}). We notice that the approaches perform comparably, with a particular distinction of Pit-all and Pit-default results which find exclusively \changing{19} and \changing{21} bugs with these high fault detection percentages instead of \changing{0}, as observed in Sub-figure~\ref{fig:venn0-plot-pit}. Nevertheless, \OurTool conserves the same advantage over the considered baselines in this regard, finding exclusively \changing{42} bugs more. It finds also \changing{50}, \changing{63} and \changing{69} more bugs than respectively Pit-all, Pit-default and Pit-rv-all, whereas they find each \changing{59}, \changing{58} and \changing{27} bugs that \OurTool missed.

\subsection{RQ3: Qualitative Analysis of {\OurTool} Mutants}
To answer this research question we investigate the mutants generated by \OurTool, which induced test suites able to find bugs that were not detected otherwise, i.e. by the considered SOA approaches (see Figure~\ref{fig:venn-plot-pit}). 
Meaning that the mutants break similar tests as the target real buggy version. 

\begin{table}[tp!]
\vspace{0.6em}
\centering
\caption{Example of mutants generated by {\OurTool} that helped find the bug Lang-49 from Defects4J.}
\begin{tabular}{l}
\toprule
Mutant 1: \textbf{replacing binary operator} \\
\midrule
\begin{lstlisting}[language=diff]
@@ org.apache.commons.lang.math.Fraction : 466 @@ 
- if (numerator == 0) {  
+ if (numerator > 0) {  
\end{lstlisting}
\\
\midrule
Mutant 2: \textbf{replacing literal implementation} \\
\midrule
\begin{lstlisting}[language=diff]
@@ org.apache.commons.lang.math.Fraction : 466 @@ 
- if (numerator == 0) {  
+ if (numerator == 1) {  
\end{lstlisting}
\\
\midrule
Mutant 3: \textbf{adding a condition to an if statement} \\
\midrule
\begin{lstlisting}[language=diff]
@@ org.apache.commons.lang.math.Fraction : 466 @@ 
- if (numerator == 0) {  
+ if ((numerator == 0) 
+   || !(numerator==Integer.MIN_VALUE)) {  
\end{lstlisting}
\\
\midrule
Mutant 4: \textbf{replacing a condition} \\
\midrule
\begin{lstlisting}[language=diff]
@@ org.apache.commons.lang.math.Fraction : 467 @@ 
- return equals(ZERO) ? this: ZERO;
+ return this;
\end{lstlisting}
\\
\midrule
Mutant 5: \textbf{replacing \texttt{this} access by another object} \\
\midrule
\begin{lstlisting}[language=diff]
@@ org.apache.commons.lang.math.Fraction : 467 @@ 
- return equals(ZERO) ? this: ZERO;
+ return equals(ZERO) ? ONE: ZERO;
\end{lstlisting}
\\
\midrule
Mutant 6: \textbf{replacing method argument} \\
\midrule
\begin{lstlisting}[language=diff]
@@ org.apache.commons.lang.math.Fraction : 469 @@ 
int gcd = greatestCommonDivisor(
- Math.abs(numerator), denominator);
+ Math.abs(numerator), 1);
\end{lstlisting}
\\
\midrule
Mutant 7: \textbf{replacing a variable} \\
\midrule
\begin{lstlisting}[language=diff]
@@ org.apache.commons.lang.math.Fraction : 473 @@ 
- return Fraction.getFraction(numerator / gcd, 
+ return Fraction.getFraction(numerator / 3, 
    denominator / gcd);
\end{lstlisting}
\\
\midrule
Mutant 8: \textbf{adding a condition to a return statement} \\
\midrule
\begin{lstlisting}[language=diff]
@@ org.apache.commons.lang.math.Fraction : 840 @@ 
return (getNumerator() == other.getNumerator() 
-  && getDenominator() == other.getDenominator());
+  && getDenominator() == other.getDenominator()))
+  || (numerator == other.numerator);
\end{lstlisting}

\\
\bottomrule
\end{tabular}
\label{tab:lang-49-mutants}
\end{table}

As a simple bug example (requiring only one change to fix it), we consider Lang-49 from Defects4J and  we investigate mutants that have been generated by \OurTool and helped in generating tests that reveal it. 
This bug impacts the results of the method \texttt{reduce()} from the class \texttt{org.apache.commons.lang.math.Fraction}, which returns a new reduced fraction instance, if possible, or the same instance, otherwise.
The bug is caused by a miss-implementation of a specific corner case, which consists of calling the method on a fraction instance that has $0$ as numerator. 
In Table~\ref{tab:lang-49-mutants}, we illustrate example mutants generated by \OurTool that helped in revealing this bug. 
Every mutant is represented by a diff between the fixed and the mutated version by {\OurTool}.

As can be seen, \OurTool can generate mutants that can be induced by applying conventional pattern-based mutations, i.e., Mutant~1 replaces a relational operator (\texttt{==}) with another (\texttt{>}) and Mutant~2 replaces an integer operand (\texttt{0}) with another one (\texttt{1}). 

In addition, it proposes more complex mutations that are unlikely achievable without any knowledge of either the AST or the context of the considered program.
For instance, 
it can generate Mutant~4 by changing a conditional return statement with (\texttt{this}) the current instance of \texttt{Fraction}, which matches the return type of the method.   
Similarly, to generate Mutant~5, it replaces (\texttt{this}) the current instance of the class \texttt{Fraction} by an existent instance of the same type (\texttt{ONE}), making the statement returning either the object \texttt{ONE} or the object \texttt{ZERO}. 

To produce more complex mutants, \OurTool applies a condition seeding followed by token-masking and CodeBERT prediction, such as adding \texttt{|| (numerator == other.numerator)} to the original condition of a \texttt{return} statement, inducing Mutant~8, or adding \texttt{|| !(numerator == Integer.MIN\_VALUE)} to the original condition of an \texttt{if} statement, inducing Mutant~3.

\begin{table*}[tp!]
\vspace{0.6em}
\centering
\caption{Example of mutants generated by {\OurTool} that helped in finding bugs from Defects4J and could not be generated when limiting the maximum number of surrounding tokens to 10.}
\begin{tabular}{l}
\toprule
Mutant 1 (JacksonCore-4) : \textbf{replacing a method call} \\
\midrule
\begin{lstlisting}[language=diff]
@@ com.fasterxml.jackson.core.util.TextBuffer : 515 @@ 
- unshare(1); 
+ expand(1);
\end{lstlisting}
\\
\midrule
Mutant 2 (Closure-26) : \textbf{replacing an object} \\
\midrule
\begin{lstlisting}[language=diff]
@@ com.google.javascript.jscomp.ProcessCommonJSModules : 89 @@ 
- .replaceAll(Pattern.quote(File.separator), MODULE_NAME_SEPARATOR)  
+ .replaceAll(Pattern.quote(filename), MODULE_NAME_SEPARATOR)  
\end{lstlisting}
\\
\midrule

Mutant 3 (Closure-35) : \textbf{replacing a method call} \\
\midrule
\begin{lstlisting}[language=diff]
@@ com.google.javascript.jscomp.TypeInference : 1092 @@ 
- scope = traverseChildren(n, scope);
+ scope = traverse(n, scope);
\end{lstlisting}
\\
\midrule
Mutant 4 (Lang-27) : \textbf{replacing a method call} \\
\midrule
\begin{lstlisting}[language=diff]
@@ org.apache.commons.lang3.math.NumberUtils : 526 @@ 
- if (!(f.isInfinite() || (f.floatValue() == 0.0F && !allZeros))) {
+ if (!(f.isInfinite() || (f.round() == 0.0F && !allZeros))) {
// also "f.floatValue()" to "f.scale()" 
\end{lstlisting}
\\
\midrule
Mutant 5 (Math-64) : \textbf{replacing an object} \\
\midrule
\begin{lstlisting}[language=diff]
@@ org.apache.commons.lang.math.Fraction : 852 @@ 
- for (int j = k; j < jacobian.length; ++j) {
+ for (int j = k; j < beta.length; ++j) {
\end{lstlisting}
\\
\midrule
Mutant 6 (Lang-27) : \textbf{replacing an object} \\
\midrule
\begin{lstlisting}[language=diff]
@@ org.apache.commons.lang3.math.NumberUtils : 526 @@ 
- if (!(f.isInfinite() || (f.floatValue() == 0.0F && !allZeros))) {
+ if (!(f.isInfinite() || (f.round() == 0.0F && !zero))) {
\end{lstlisting}

\\
\bottomrule
\end{tabular}
\label{tab:10tokens-mutants}
\end{table*}

To investigate further the impact of the code context captured by the model on the generated mutants, we have rerun \OurTool on 5 subjects from our dataset, with a maximum number of surrounding tokens equal to 10 (instead of 512). 
Then, we compared manually the induced mutants with those generated by our default setup, in the same locations.
From our results, we observed a noticeable decrease in the number of compilable predictions, indicating the general performance decrease of the model when it lacks information about the code context.
Particularly, we notice that it is not able to produce program-specific mutants, i.e. by changing an object by another or a method call with another. 
In Table~\ref{tab:10tokens-mutants}, we illustrate some example mutants that helped find each of the studied subjects (breaking same tests as the original bug), which \OurTool failed to generate when the maximum number of surrounding tokens is limited to 10.

\section{Threats to Validity}
\label{sec:threats-to-validity}
One external threat to validity concerns the generalisation of our findings and results in the empirical evaluation. 
To reduce this threat, we used a large number of real bugs from popular open-source projects with their associated developer test-suites, provided by an established and independently built benchmark (i.e. Defects4J~\cite{defects4j}). 
Nevertheless, we acknowledge that the results may be different considering projects in different domains.

Other threats may arise from our way of assessing the fault detection capability of mutation testing approaches, based on their capability of guiding the testing via a developer/tester simulation in which we assume that the current test suites are complete and the not killed mutants are equivalent. 
Although we acknowledge that this may not be the case in real-world scenarios, we believe that this process is sufficient to evaluate our approach, particularly considering the fact the test suites provided by Defects4J are relatively strong. 
Additionally, to mitigate any comparison threat between the considered approaches, we use consistently and similarly the same test-suites, setups and simulation assumptions in all our study. 

The choice of our comparison baseline may form other threats to the validity of our findings. While different fault-seeding approaches have been proposed recently, PiTest remains among the most mature and stable mutation testing tools for Java programs, thus, forming an appropriate comparison baseline to evaluate our work. 
Furthermore, we compared our results with those obtained by mutants from different configurations proposed by PiTest, enlarging our study to the different audiences targeted by this latter.
We acknowledge however that the results may change when considering other techniques and consider the evaluation of the effectiveness and cost-efficiency of different mutation testing techniques as out of the scope of this paper.

Other construct threats may arise from considering the number of mutants analysed as metric to measure the effort required to find a fault. 
In addition to the fact that this metric has been widely used by the literature \cite{PapadakisK00TH19,AndrewsBLN06,KurtzAODKG16}, we believe that it is intuitive and representative of the global manual effort of the tester in analysing the mutants, discarding them or writing tests to kill them.
While being the standard in the literature, we acknowledge that this measure does not account for the cost difference between mutants, attributing the same cost to all mutants. This is simply because we do not know the specific effort required to analyse every specific mutant or to write every specific test.
Additionally, our cost-efficiency results may be impacted by costs that are not captured with this metric, such as the execution or the developing effort of either CodeBERT, the key component of \OurTool, or the set of patterns and execution enhancements over the different releases of PiTest. 
Nevertheless, we tried to mitigate any major threats that can be induced by the implementation of the different tools, i.e. we reduce the dataset subjects to those on which every approach generated at least one mutant and consider any implementation difference between the approaches as out of the current scope.

\section{Related Work}

Since the 1970s, mutation testing has been the main focus of multiple research works~\cite{10.5555/275587}. 
Their findings have proven that artificial faults can be useful in multiple software engineering applications, such as testing \cite{PapadakisK00TH19}, debugging~\cite{papadakis2015metallaxis,LouGLZZHZ20}, assessing fault tolerance \cite{NatellaCDM13}, risk analysis \cite{ChristmanssonC96, VoasCMMF97} and dependability evaluation \cite{DArlatCCLP93}.

Despite this long history of research, the generation of relevant mutants remains an open question. 
Most of the related research has focused on the design of fault patterns (mutation operators) which are usually defined based on the target language grammar~\cite{0020331, PapadakisK00TH19} then refined through empirical studies~\cite{OffuttLRUZ96, MarcozziBKPPC18, KintisPPVMT18} aiming at reducing the redundancy and noise among their generated mutants. 
The continuous advances in this sense were followed by a constant emergence of pattern-based mutation testing tools and releases~\cite{mujava,major,pitest}, among which some are becoming popular and widely adopted by researchers and practitioners, such as PiTest~\cite{pitest}, from which we consider three configurations as our comparison baseline.

Recent research has focused their interest on improving the representativeness of artificial faults aiming at reducing the mutation space to real-like faults. 
For instance, instead of basing the mutation operators' design on the programming language grammar, Brown et al.~\cite{BrownVLR17} proposed inferring them from real bug fixes. 
Similarly, Tufano et al.~\cite{TufanoWBPWP19} proposed a neural machine translation technique that learns how to inject faults from real bug fixes. 
Along the same line, Patra et al.~\cite{SemSeed} proposed a semantic-aware learning approach, that learns and then adapts fault patterns to the project of interest.  
Their results are promising, however, the fact that these techniques depend on the availability of numerous, diverse, comprehensive and untangled fix commits~\cite{herzig2011untangling} of not coupled faults~\cite{Offutt92}, which is often hard to fulfil in practice, may hinder their performance.
Acknowledging for the injection location~\cite{NatellaCDM13,ChekamPBTS20}, 
Khanfir et al.~\cite{khanfir2020ibir} combined the usage of information retrieved from bug reports with inverted automated-program-repair patterns to replicate real faults fixable by the original fix-patterns. 
Their results showed that they can generate faults that mimic real ones, however, their approach remains dependent and limited to the presence of good bug reports.  
Overall, designing the mutation operators based on the known faults space yields more diverse mutants that represent more fault types. 
However, these extended operator sets tend to increase the number of generated mutants and consequently the general cost of the mutation campaign i.e. the fault patterns proposed by Brown et al. and Khanfir et al. counted also most of the conventional mutators in addition to new ones. 
Unlike these techniques, \OurTool leverages pre-trained models to introduce mutants based on code knowledge instead of the faults one. 
As code is more available than faults, it offers a more flexible and complete knowledge base than faults, i.e. it perms to overcome the limitations and efforts required 1) to collect clean bug-fixing commits, 2) to capture the faulty behaviour and 3) design fault patterns, be it manually or via machine learning techniques. 

Aiming at reducing the number of generated mutants, researchers have proposed different strategies to generate relevant mutants. 
For instance, studies that show that mutant strength resides in not only its inducing pattern but also where it is injected~\cite{NatellaCDM13,ChekamPBTS20}, motivated the importance of selecting relevant locations to mutate. 
In this regard, Sun et al.~\cite{SunXLZ17} suggest mutating multiple places within diverse program execution paths. Gong et al.~\cite{GongZYM17} also propose the mutation in diverse locations of the program extracted from graph analysis. Similarly, Mirshokraie et al.~\cite{Mirshokraie0P15} compute complexity metrics from program executions to extract locations with good observability to mutate.
Other approaches restrict the fault injection on specific locations of the program, such as the code impacted by the last commits~\cite{ReMT,relevantMutantsICSME2020} for better usability in continuous integration, or targeting locations related to a given bug-report~\cite{khanfir2020ibir} to target a specific feature or behaviour, etc. 
More recent advances have resulted in powerful techniques for cost-effectively selecting mutants, i.e., by avoiding the analysis of redundant mutants (basically, equivalent and subsumed ones)~\cite{GODCPT, JuniorDDSVD20, GheyiRSGFdATF21}. In particular, the work of Garg et al.~\cite{GODCPT} utilises the knowledge of mutants' surrounding context, embedded into the vector space, to predict whether a mutant is likely subsuming or not. 
In this work, we do not target any specific code part or any narrow use case, but instead, perform fault injection in a brute-force way similarly to mutation testing, by iterating every program statement and masking every involved token.  


Multiple studies have been focused on the relationship between artificial and real faults~\cite{PapadakisK00TH19}. 
The results of the studies conducted by Ojdanic et al.~\cite{ojdanic2021syntactic}, Papadakis et al.~\cite{PapadakisSYB18}, Just et al.~\cite{JustJIEHF14} and Andrews et al.~\cite{AndrewsBLN06} showed that there is a correlation between tests broken by a bug and tests killing mutants. Meaning that artificial faults can be used as alternatives to real faults in controlled studies. 
Moreover, the findings of Chekam et al. \cite{ChekamPTH17}, Frankl et al. \cite{FranklWH97} and Li et al. \cite{LiPO09} show that guiding testing by mutants leads to significantly higher fault revelation capability than the ones of other test adequacy criteria. 
Based on these findings, we assess our approach based on the relation between the injected and real faults, in terms of breaking tests. 
More precisely, we conduct a fault detection effectiveness and cost-efficiency study to evaluate our approach's mutants in guiding testing and compare it to state-of-the-art techniques. 
Furthermore, we discuss the diversity and readability of \OurTool mutants through real examples.

The closest related work is a preliminary implementation of {\OurTool} that was recently presented in the 2022 mutation workshop~\cite{DBLP:conf/icst/DegiovanniP22}. 
This implementation, denoted as {\Toolconv} in our evaluation, includes the conventional mutations (to mask and replace tokens by the model predictiosn), but it does not include the condition-seeding additive mutations that provide major benefits for fault detection. 
Moreover, {\Toolconv} was evaluated only on 40 bugs from Defects4J, and compared only to an early version of PiTest (similar to Pit-rv-all). 
In this work, we perform an extensive experimental evaluation including 689 bugs from Defects4J and compare {\OurTool} effectiveness with three different configurations from PiTest. Moreover, we show that {\OurTool} finds on average more bugs than {\Toolconv} without requiring more effort.

\section{Conclusion}

We presented \OurTool; a pre-trained language model based fault injection approach. 
\OurTool provides researchers and practitioners with easy-to-understand ``natural'' mutants
to help them in writing tests of higher fault revelation capabilities. 

Unlike state-of-the-art approaches, it does neither require nor depend on any kind of faults knowledge or language grammar but instead on the actual code definition and distribution, as written by developers in numerous projects.
This facilitates its developing, maintainability, integration and extension to different programming languages. In fact, it reduces the overhead of learning how to mutate, be it via creating and selecting patterns or collecting good bug-fixes and learning from their patches.

In a nutshell, \OurTool takes as input a given program and replaces different pieces of its code base with predictions made by a pretrained generative language model, producing multiple likely-to-occur mutations. 
The approach targets diverse business code locations and injects either simple one-token replacement mutants or more complex ones by extending the control-flow conditions. 
This provides probable developer-like faults impacting different functionalities of the program with higher relevance and lower cost to developers.
This is further endorsed by our results where \OurTool induces high fault detection test suites at low effort, outperforming state-of-the-art techniques (PiTest), in this regard. 

We have made our implementation and results available~\cite{mBERT-nt} to enable reproducibility and support future research.



\section*{Acknowledgment}
This work was supported by the Luxembourg National Research Fund (FNR) projects C20/IS/14761415/TestFlakes and TestFast, ref. 12630949.

\balance
\bibliographystyle{plain}
\bibliography{bibfile}

\end{document}